\documentclass[10pt, conference]{IEEEtran}
\usepackage[utf8]{inputenc}
\usepackage{multicol}
\usepackage[margin=0.75in]{geometry}
\usepackage{authblk}
\usepackage[numbers,sort&compress,square]{natbib}
\usepackage{graphicx}
\usepackage{amsmath}
\usepackage{url}
\usepackage{caption}
\usepackage{subcaption}
\usepackage{xcolor}
\usepackage{algorithm}
\usepackage {algpseudocode}
\usepackage[justification=centering]{caption}
\begin{document}

\title{Cellular Network Capacity and Coverage Enhancement with MDT Data and Deep Reinforcement Learning}

\author[1, *]{Marco Skocaj}
\author[1]{L.M. Amorosa}
\author[2]{G. Ghinamo}
\author[2]{G. Muratore}
\author[2]{D. Micheli}
\author[1]{F. Zabini}
\author[1]{R. Verdone}
\affil[1]{DEI, University of Bologna \& WiLab - National Laboratory for Wireless Communications (CNIT)}
\affil[2]{TIM Telecom Italia}
\affil[*]{\textit{Corresponding author: marco.skocaj@unibo.it}}

\renewcommand\Authands{ and }

\date{21st February 2022}

\maketitle

\begin{abstract}
\noindent \textbf{Recent years witnessed a remarkable increase in the availability of data and computing resources in communication networks. This contributed to the rise of data-driven over model-driven algorithms for network automation. This paper investigates a Minimization of Drive Tests (MDT)-driven Deep Reinforcement Learning (DRL) algorithm to optimize coverage and capacity by tuning antennas tilts on a cluster of cells from TIM’s cellular network. We jointly utilize MDT data, electromagnetic simulations, and network Key Performance indicators (KPIs) to define a simulated network environment for the training of a Deep Q-Network (DQN) agent. Some tweaks have been introduced to the classical DQN formulation to improve the agent’s sample efficiency, stability and performance. In particular, a custom exploration policy is designed to introduce soft constraints at training time. Results show that the proposed algorithm outperforms baseline approaches like DQN and best-first search in terms of long-term reward and sample efficiency. Our results indicate that MDT-driven approaches constitute a valuable tool for autonomous coverage and capacity optimization of mobile radio networks.}
\end{abstract}

\section{Introduction}
As next-generations network management is foreseen to reach a level of complexity beyond human's ability to fully comprehend and control \cite{Ericsson_WP_AI_nextgen} , the last decade has been characterized by an up-growing interest in network automation-related services and applications. The requirement to meet the needs of both users and operators in a cost effective way has triggered research to add intelligence and autonomous adaptivity into future cellular networks \cite{survey_SON}. The first network automation technology introduced by 3GPP was Self-Organizing Network (SON), initially designed for Long Term Evolution (LTE) technology \cite{3GPP_36.902, 3GPP_32.500}. SON aims to reduce manual involvement in planning, configuration, management, optimization, and healing of mobile radio access networks. The next step towards the minimization of human intervention is the zero-touch paradigm, for which pervasive artificial intelligence (AI) algorithms will play a fundamental role. Recent years also witnessed an unprecedented availability of data and computing resources in many engineering domains \cite{Simeone}. This contributed to the rise of data-driven over model-driven AI. The benefit of data-driven approaches for next-generation networks, particularly for coverage and capacity optimization (CCO) tasks, has already been discussed in the literature \cite{Big_Data_Networks}.\\
Minimization of Drive Tests (MDT) is a feature introduced in 3GPP Release 10 \cite{3GPP_37.320} that enables operators to utilize users' equipment (UE) to collect radio measurements and associated location information \cite{MDT} (e.g., via GPS or WLAN localization (Release 16)). Collected MDT measurement reports can be aggregated into datasets with user location information (ULI) as well as statistical channel state information, providing a realistic description of the network state and allowing network troubleshooting and inference of human behavior-related statistics \cite{MDT_Venezia, MDT_Nokia_TIM, MDT_concert}. With this contribution, the authors aim to shed light on the potential of MDT data-driven algorithms for network automation. To this end, a CCO problem is addressed for a selected cluster of the TIM Italy's network by optimizing cells' antenna downtilts with the use of an MDT-driven Deep Reinforcement Learning (DRL) algorithm.

\subsection{State of the art}

CCO is a challenging optimization problem of cellular networks which has widely been studied in the literature, especially in the broader subject of SON \cite{survey_SON, CCO_SON, CCO_SON2}. One way to tackle CCO is to allow coverage regions to change by modifying cells' elevation tilts. Some early decade contributions made use of research operations methods \cite{CCO_SON, CCO_SON2} to tune antennas parameters, while most recent years, motivated by the late success of Deep Learning (DL) and Reinforcement Learning (RL), have seen the rise of solutions employing DRL algorithms \cite{Andrews, R.W.Heath, RET_safe_RL, Yueming}. RL, indeed, seems to be a valuable tool for CCO since it can learn and adapt to the dynamics of the environment \cite{Andrews}. In \cite{Andrews}, a two-step algorithm (multi-agent mean-field RL and single-agent RL) provides a scalable solution for online antenna tuning of a multi-tier network. In \cite{R.W.Heath}, the CCO problem is addressed by balancing under-vs-over coverage using Deep Deterministic Policy Gradient (DDPG) and Bayesian Optimization. In \cite{Yueming}, a solution for optimizing the antenna parameters of a HetNet is proposed addressing the unstable convergence of hyperparameters. Finally, \cite{RET_safe_RL} proposes a safe antenna tilt update policy to avoid the execution of degrading actions and improve system's reliability. All of the aforementioned papers are innovative, as they respectively propose solutions to improve scalability \cite{Andrews, Yueming}, sample-efficiency \cite{R.W.Heath} and reliability \cite{RET_safe_RL}. Nevertheless, none of them makes use of realistic data measurements like MDT for network state representation and training. At most, \cite{R.W.Heath} employs electromagnetic simulations for Reference Signal Received Power (RSRP) map generation. In this paper, the agent is trained by direct interaction with a simulated network environment jointly employing MDT data, network KPIs, and electromagnetic simulations from the TIM cluster.

\subsection{Paper contributions}

\noindent The proposed work aims to show the potential of MDT data-driven algorithms for network automation. To the best of the authors' knowledge, the joint use of MDT data and DRL constitutes a novel approach to CCO problems. Even though being standardized by 3GPP as a key SON feature \cite{3GPP_37.320}, MDT data have not been extensively investigated in the literature of automated communication systems. Moreover, their application is mainly circumscribed to anomalous coverage detection problems \cite{MDT_CD_problems, MDT_CD_problems_2, MDT_CD_problems_4} and Quality of Service (QoS) verification \cite{MDT_CD_problems_3}. The only contribution where MDT data are employed in the context of a capacity optimization problem is \cite{COM_HetNets}, where a two-steps algorithm - cell outage detection and cell outage compensation - is proposed for the management of a split-architecture network.\\
Nevertheless, only the former employs MDT-based anomaly detection to identify coverage issues, while the latter makes use of an actor-critic RL algorithm based on control/data signals to adjust antenna settings. It follows that, even in this contribution, MDT data are relegated to coverage issues detection. The benefits of employing a dataset with radio measurement reports and ULI for capacity optimization, instead, are manifolds and discussed throughout the following sections.\\

\noindent Another element of distinction with the literature concerns the approach adopted for the optimizer. The use of Deep-Q Networks (DQN) as Q-function \(Q_\Theta(s, a)\) approximators is known and widely utilized in literature \cite{human_level_control, atari_DRL}, but is affected by sample inefficiency since the DQN agent struggles to explore the states' space efficiently and exhaustively. Consequently, some of the works in the literature employed other sample-efficient algorithms for CCO problems.  In \cite{R.W.Heath}, the authors compare the sets of Pareto frontiers for balancing under-vs-over coverage found by using DDPG and Bayesian Optimization (BO), with the result of DDPG obtaining slightly better performance at the expense of two orders of magnitude slower exploration phase. Instead, the effort of this work is focused on identifying a method for improving a centralized DQN agent's sample efficiency. To this end, we introduced some tweaks, thoroughly described in Sec. 5, to the classical DQN formulation. In particular, a custom exploration policy, namely depth-wise-\(\epsilon\eta\)-greedy (DW-\(\epsilon\eta\)-greedy) policy is proposed to introduce soft constraints at training time while adequately balancing the exploration vs. exploitation tradeoff. Results show that the agent's efficiency improved both in terms of training time, stability and performance. Thanks to these enhancements, it was possible to fit the antenna tuning problem to a medium-sized cluster covering a geographic area of approximately 26 [$Km^2$].\\

\noindent Multi-agent systems solutions are currently capturing significant research interest because they can remarkably enhance the scalability of network optimization algorithms \cite{RRM_MADRL, Andrews, MADRL_power_WN, MF_MARL, MEK_MARL, DMA_MEC}. The design of distributed multi-agent systems is sometimes also a necessity motivated by the partial state observability of edge network elements. Nonetheless, such design poses many challenges, like efficient communication between agents \cite{learning_to_communicate, Gunduz, MG_BA_MADRL} and non-stationarity \cite{Review_MARL}, oftentimes leading to unstable training and sub-optimal solutions. Be that as it may, numerous reasons motivated a centralized approach for this work:
\begin{itemize}
\item The use of MDT data allows full state observability with respect to other kinds of signals and data used to represent the observation space in RL-communications problems.
\item MDT data are collected from TIM's network infrastructure, therefore we must abide to actual cell deployment. Given the current network density, a CCO problem can be effectively tackled with a single agent approach even for urban areas of medium-sized cities like Bologna, Italy.
\item SON envisages three kinds of architectures: centralized, hybrid, distributed. A centralized solution is preferred for optimality and stationarity, whereas a distributed one is imperative to tackle the curse of dimensionality. The best tradeoff approach for larger cities or denser urban areas might envision a hybrid solution where multiple agents are responsible for variable-sized adjacent clusters instead of single e/gNodeBs. These hybrid agents might learn to maximize a common reward cooperatively based on their mild-to-severe mutual interference. To this end, the proposed contribution is also propaedeutic for future work on distributed/hybrid solutions for larger or denser geographical areas. Previous works in the literature already demonstrate that DQNs can become a practical tool for studying the decentralized learning of multiagent systems living in highly complex environments \cite{Cooperation_Competition_in_MARL}.\\
\end{itemize}

\noindent The rest of the paper is organized as follows: Sec. 2 describes the System model. Sec. 3 provides a detailed description of the simulated network environment. Sec. 4 formalizes the problem as Markov Decision Process (MDP), whereas Sec. 5 illustrates the details introduced for tweaking a DQN agent. Finally, Sec. 6 presents the obtained results and analyzes the agent's compactness and stability, while Sec. 7 draws the final remarks.

\section{System Model}

\begin{figure}[ht]	
\includegraphics[width=\columnwidth]{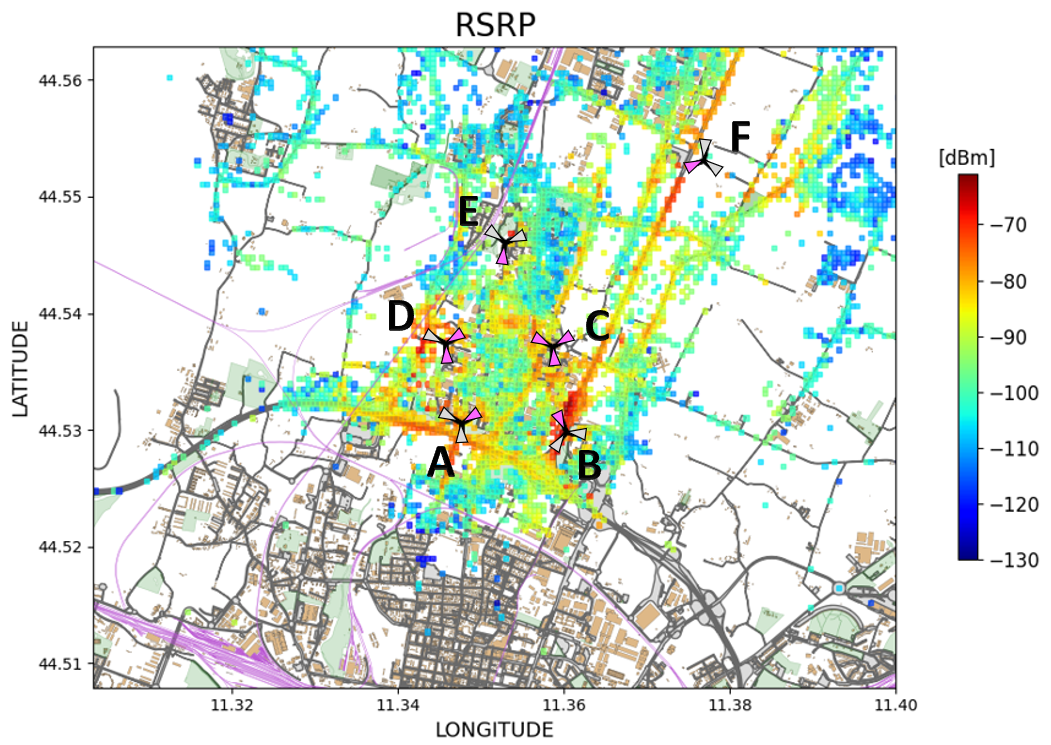}
\centering
\caption{TIM's Network deployment - North Bologna Area - RSRP measurements from MDT}
\label{fig: Corticella Area}
\end{figure}

The system model comprises a cellular network deployment for a selected area of interest, incorporating a total number of "target" and "boundary" eNodeBs. The former are the ones actively targeted by a DRL agent, whereas the latter are out of the scope of the optimization process but are still taken into account to consider boundary effects of the agent's actions. In particular, the reference scenario involves TIM's network LTE deployment for an area of approximately 26 [$Km^2$] in the North of the city of Bologna, Italy (Fig. \ref{fig: Corticella Area}), incorporating a total of 18 eNodeBs (6 sites with tri-sectorial distribution). Among these, 9 eNodeBs have been selected as "target nodes" (pink markers, Fig. \ref{fig: Corticella Area}), while the remaining ones (grey markers) are selected as "boundary" cells. The selected area is interesting and challenging from the network configuration point of view since it yields a heterogeneous propagation scenario: highway, urban areas, and agricultural fields.\\
\\
\noindent A Reinforcement Learning agent for network parameters optimization could be trained by direct interaction with a physical network. However, this is generally very inconvenient due to slow training times and potential degrading actions performed during the exploration phase of the training. In this regard, many contributions in literature focused on risk sensitivity and damage avoidance during the exploration phase of an agent under the mantle of safe RL \cite{safe_RL,RET_safe_RL}.\\

\noindent Nevertheless, safe RL is not addressed by this work since, due to the following reasons, an offline approach was chosen for the training of the agent:
\begin{enumerate}
\item The simulated network environment, as described in Sec. 3, is based on MDT data processing. The possibility of relying on time-referenced and geo-located actual traffic distributions, as well as measured radio indicators, provides a far more accurate insight of the network with respect to solutions based on simulated traffic and model-based received power. For instance, measured RSRP and Reference Signal Received Quality (RSRQ) intrinsically hold local information such as LoS/NLoS UE's condition, material and shape of surrounding buildings, UE's specific antenna model, average human-body absorption, and many more.
\item For an online training, a state space based on real-time reward signal should be employed. Consequently, it would be impossible to utilize MDT data, with the disadvantage of losing ULI.
\item Antenna tuning's dynamic is highly correlated to changes in traffic patterns. Such changes evolve at two different paces:
\begin{itemize}
\item \textit{Fast}: different antenna configurations might be well-suited for different day periods (e.g., the flow of people leaving their offices and populating roads) but occur with a daily and weekly periodicity, as evident from Fig. 2c and 2a.
\item \textit{Slow}: long-term differences in daily traffic patterns are observable over extensive periods, like multiple weeks or months. Fig. \ref{fig:figures}a and \ref{fig:figures}b show autocorrelation plots of the normalized number of RRC connected UEs for the city of Bologna over few weeks (Fig. 2a) and a full year (Fig. 2c). The monotonically decreasing behavior of the autocorrelation plot denotes an increasing variability in the daily traffic patterns.
\end{itemize}

\begin{figure*}[t]

\centering
\begin{subfigure}{0.3\textwidth}
    \includegraphics[width=\textwidth]{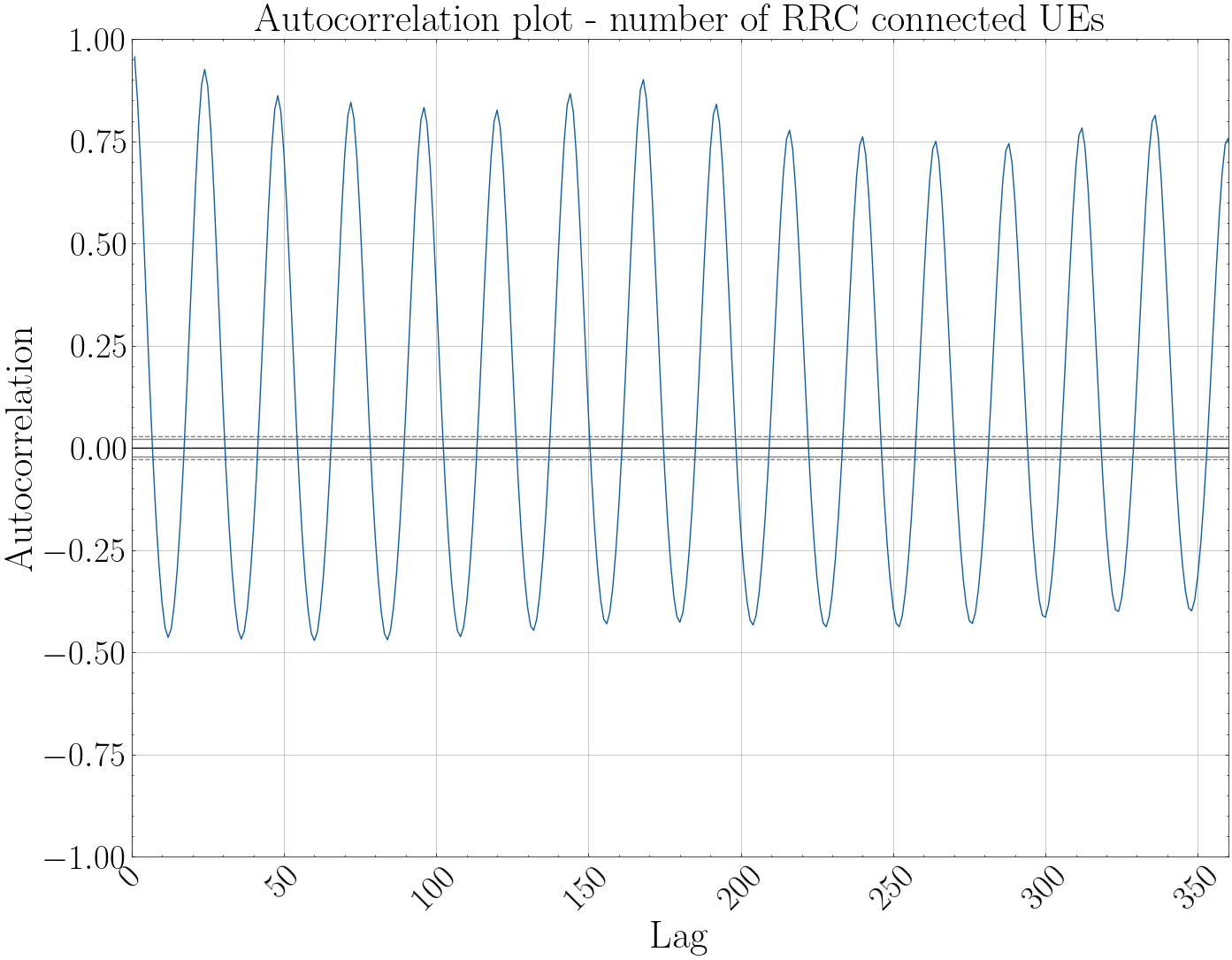}
    \caption{RRC Connected UEs' autocorrelation plot - 2 weeks}
    \label{fig:RRC_2w}
\end{subfigure}
\hfill
\begin{subfigure}{0.3\textwidth}
    \includegraphics[width=\textwidth]{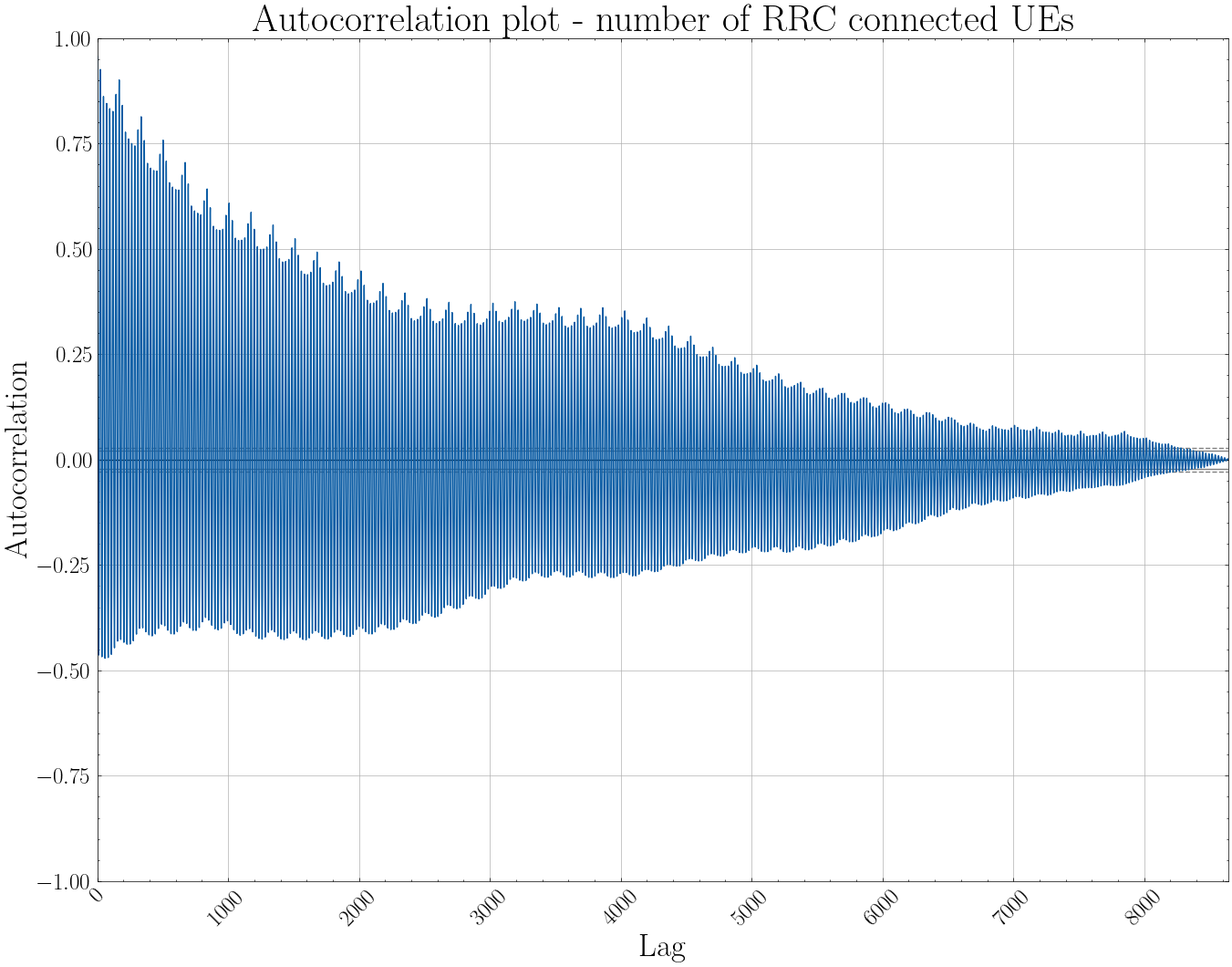}
    \caption{RRC Connected UEs' autocorrelation plot - 1 year}
    \label{fig:RRC_1y}
\end{subfigure}
\hfill
\begin{subfigure}{0.3\textwidth}
    \includegraphics[width=\textwidth]{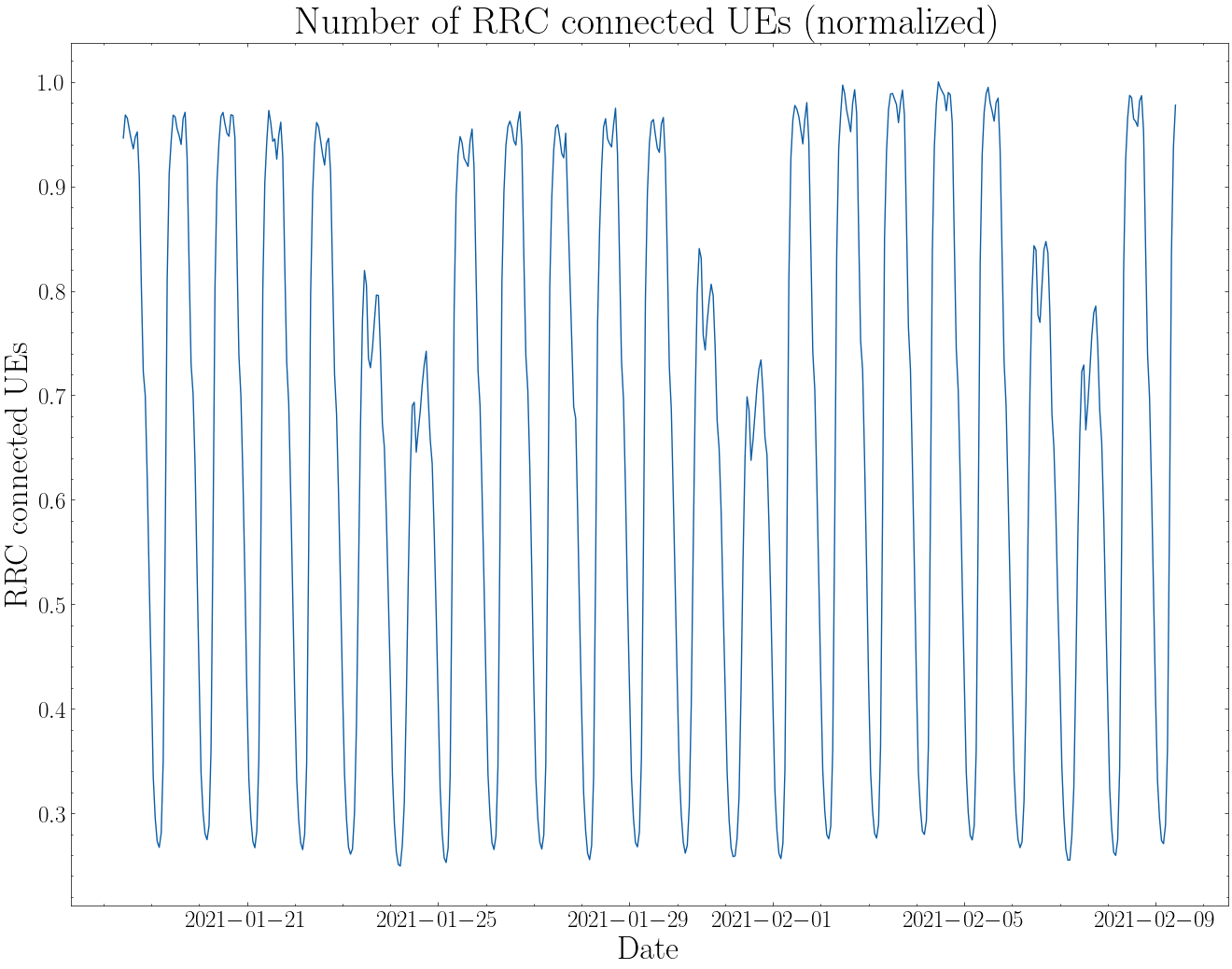}
    \caption{Number of RRC Connected UEs (normalized)}
    \label{fig:RRC_norm}
\end{subfigure}
\centering
\caption{Daily periodicity is evident from Fig. (a) and (c). Weekly periodicity is observable from a slight increase of the autocorrelation values over a 7 day period Fig. (a).}
\label{fig:figures}

\end{figure*}

Parallel training of specialized agents might address the problem of fast dynamicity for different time ranges manifesting quasi-static behaviors and exploiting patterns' daily periodicity. The agents' policies validity would last for a much larger period than training time. Considering a scenario where trained agents are employed and running on a physical network, MDT measurement campaigns might be periodically collected to test agents' performance and proactively identify outdated optimizers.\\
\end{enumerate}

\begin{figure}[h]	
\includegraphics[width=\columnwidth]{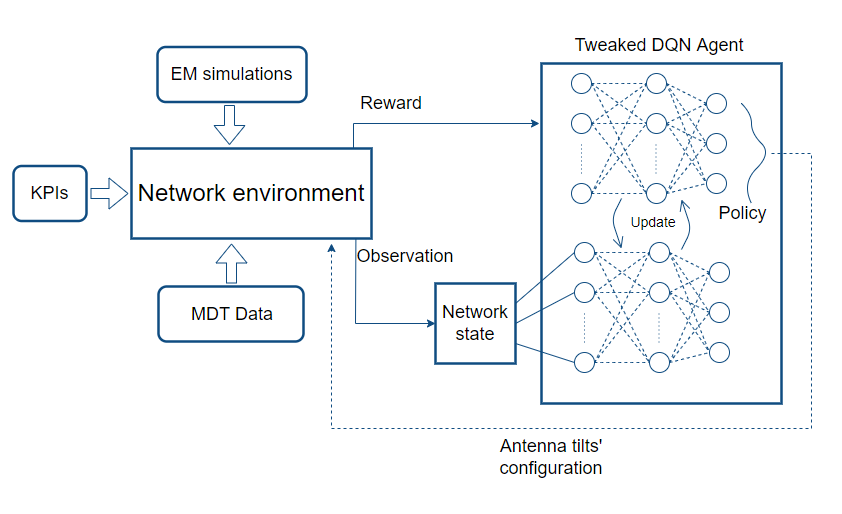}
\centering
\caption{System Model}
\label{fig: System Model}
\end{figure}

\noindent The overall system framework is well-suited for a block description (Fig. \ref{fig: System Model}).

\noindent On the left side, a simulated network environment receives MDT data, traffic KPIs, and electromagnetic simulations as inputs. The scope of the environment is to accurately represent the effect that the reconfiguration of a parameter (e.g., eNodeBs' antenna tilts), which corresponds to an agent's action, has on network performance. To this end, the environment involves a set of data processing methods, accurately described in Sec. 3. Electromagnetic simulations have been generated considering single eNodeB's antenna tilt configurations (a total of $N_{cells} \times N_{tilts}$ simulations) with the use of a TIM's proprietary tool. In contrast, MDT data and KPIs have been collected from multiple-days measurements. The simulated environment interacts with a tweaked DQN agent during training time, described in Sec. 5. The agent's goal is to eventually learn which actions would lead to a maximized cumulative future reward \(R_c\) based on its environment observation. The latter can be expressed as:
\begin{equation}
\centering
\label{eqn:(1)}
R_c = \sum_{i=t}^{\infty}\gamma^{i-t} R_{\pi(s_i)}(s_i,s_{i+1})
\end{equation}
where $\gamma$ is a discount factor for future rewards and $R_{\pi(s_i)}$ is the immediate reward obtained after the agent's policy \(\pi\) selects action \(a_i\) based on observation of state \(s_i\).\\

\noindent As better described in the next two sections, the quantities in Tab. \ref{Tab:Tabella1} have been collected from MDT measurement reports to build the observation space for the agent.\\

\begin{table}[ht]
\centering
\resizebox{\columnwidth}{!}{%
  \begin{tabular}{|l|l|}
  \hline
    PCELL\_RSRP & \textit{UE's measured RSRP from the serving primary cell}\\
    \hline
    NCELLS\_RSRP & \textit{UE's measured RSRP from up to 8 adjacent non serving cells} \\
    \hline
    PCELL\_RSRQ & \textit{UE's measured RSRQ from the serving primary cell}\\
    \hline
    CALL\_IDs & \textit{Anonymous temporary ID identifying a UE's RRC connection}\\
    \hline
  \end{tabular}%
  }
  \centering
  \caption{MDT observation space}
  \label{Tab:Tabella1}
\end{table}

\noindent Moreover, two traffic KPIs were gathered with a cell-level spatial granularity: \textit{average active number of UEs per Time Transmission Interval}, from now on referred to as \textit{act\_UEs}, and \textit{average cell load} \(\rho\).

\section{Simulated Network Environment}

\begin{figure}[h]
\includegraphics[width=\columnwidth]{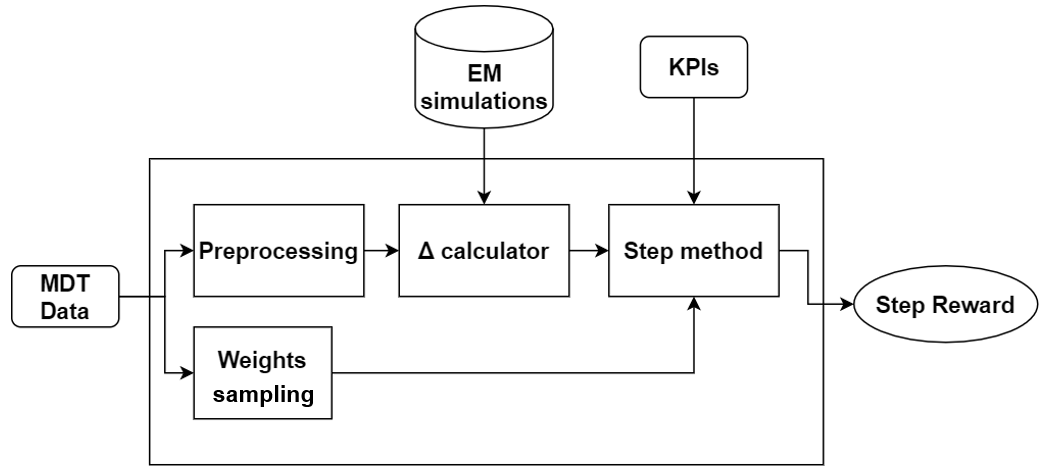}
\centering
\caption{Simulated environment - block system view}
\label{fig: env_blocks}
\end{figure}

\noindent The simulated network environment is composed by 4 logical blocks performing different tasks, summarized by Fig. \ref{fig: env_blocks}.

\subsection{Pre-processing}

\noindent The collected MDT data are divided into time slots exhibiting quasi-static traffic behavior (e.g., 2 P.M. - 6 P.M.), determined through the \textit{act\_UEs} KPI. MDT data are then aggregated into pixels of 50 \([m^2]\) dimension, aligned coherently with the ones produced by an electromagnetic simulation tool. \cite{enhanced_mdt-based} describes how the optimal area division into bins (pixels) affects three kinds of MDT measurement reports errors: 1) Positioning error, 2) Quantization error, 3) Scarcity of reports. The pixel size in this contribution has been determined prioritizing the minimization of errors 1 and 3, since a quantization error due to averaging over larger bins does not significantly affect the agent’s decision on antenna tilt parameters.

Only pixels with statistical relevance are retained. Statistical relevance is determined based on a twofold condition:
\begin{enumerate}
\item The number of MDT samples in every pixel exceeds a pre-defined threshold. To this end, the minimization of the scarcity of reports error translates to a lower \% of discarded pixels. 
\item At least one MDT sample in every pixel yields measurements of a minimum of three interfering cells.
\end{enumerate}
Another motivation behind the pixelization process of MDT data is to reduce the variance of the single measurements and provide a dimensional-consistent input to the DQN. Each pixel is fully characterized by three aggregated quantities that are thus computed:\\
\begin{enumerate}
\item \textbf{RSRP}: linear average of RSRP measurements, transformed in dB.\\
\item \textbf{WEIGHT}: The default weight is determined as the number of different CALL\_IDs of a pixel in the reference time slot. As previously expressed, a CALL\_ID relates to a single RRC\_Connection establishment. For Immediate mode MDT, RRC\_Connected UEs are configured for periodic measurement reporting (in our setup, every 5 seconds). Consequently, a very different number of reports might be produced according to each UE’s connection time. Hence, assigning weights based on the number of connections over the total number of samples prioritizes fairness among users.
At the beginning of a new episode, each pixel weight is sampled from a Poisson probability distribution (Weights sampling block, Fig. \ref{fig: env_blocks}), in which the average value \(\lambda\) is set equal to the number of original CALL\_IDs per pixel. This procedure confers generality to the environment and prevents the agent from overfitting the training data.\\
\item \textbf{SINR}: Signal-to-Noise and Interference Ratio (SINR) computation can be carried out following either one of the two proposed formulas:
\begin{enumerate}
\item \textit{RSRP-based SINR}:
\begin{align}
\label{eqn:(2)}
&\text{SINR}[dB]=\text{RSRP}_{\text{Pcell}}[\text{dBm}] \nonumber \\ &-10\log_{10}\left( \sum_{i=1}^{N}\text{RSRP}_{\text{Ncell},i}[\text{mW}]\right. \nonumber \\
&\left. +N_0\left[\frac{\text{mW}}{\text{Hz}}\right]\cdot B_{R}[\text{Hz}]\right)
\end{align}
At the numerator, the average measured RSRP from the pixel's serving cell, while, at the denominator, interference is expressed as a sum of interfering cells' RSRP. Noise is accounted assuming a noise figure of 7 [dB] and reference and antenna temperature equal to 290 [K]. The noise power referred to a RE bandwidth of 15 [kHz], according to RSRP definition \cite{3GPP_36.214}, is therefore approximately -125[dBm].\\
\item \textit{RSRQ-based SINR}:
\begin{equation}
\label{eqn:(3)}
\centering
\text{SINR}=\frac{12 \cdot \text{RSRQ}}{1-\rho \cdot \text{RSRQ} \cdot 12}
\end{equation}
A second possible method for evaluating SINR starts from the RSRQ and {average cell load} \(\rho\), indicating the \% of occupied Physical Resource Blocks (PRBs) in the corresponding serving cell. 12 is the number of OFDM sub-carriers in an LTE PRB, RSRQ at the numerator is the RSRQ aggregated on an MDT data pixel basis, in linear scale. RSRQ is defined as \(\frac{N_{PRB}\cdot \text{RSRP}}{\text{RSSI}}\) \cite{3GPP_36.214}, where \(N_{PRB}\) is the number of resource blocks of the E-UTRA carrier Reference Signal Strength Indicator (RSSI) measurement bandwidth. Since RSRP is a measurement of the demodulated Channel Reference Signal (CRS) symbols, the numerator accounts only for the co-channel serving cell signal contribution. In contrast, the denominator's RSSI is a wideband measure of co-channel serving and nonserving cells, adjacent channel interference, and noise. As a consequence, (\ref{eqn:(3)}) is a better descriptor of the SINR when performance are limited by interference and susceptible to load variations in the cells, compared to (\ref{eqn:(2)}), where SINR is evaluated as a ratio of reference signals demodulated powers.
\end{enumerate}
\end{enumerate}

\begin{figure*}[h]
\centering
\begin{subfigure}{0.4\textwidth}
    \includegraphics[width=\textwidth]{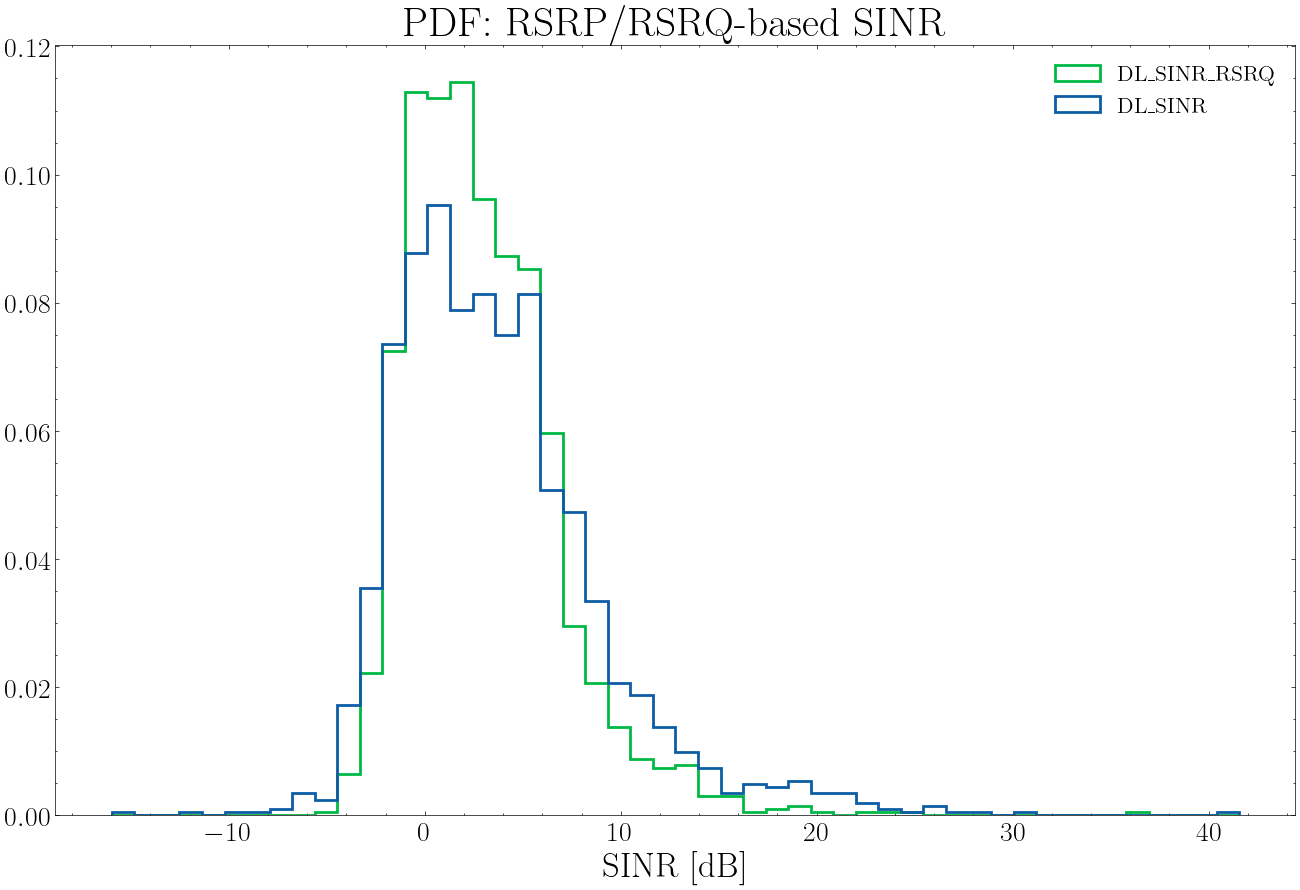}
    \caption{PDF: RSRP/RSRQ-based SINR}
    \label{fig:subfig_RSRP/Qa}
\end{subfigure}
\hfill
\begin{subfigure}{0.4\textwidth}
    \includegraphics[width=\textwidth]{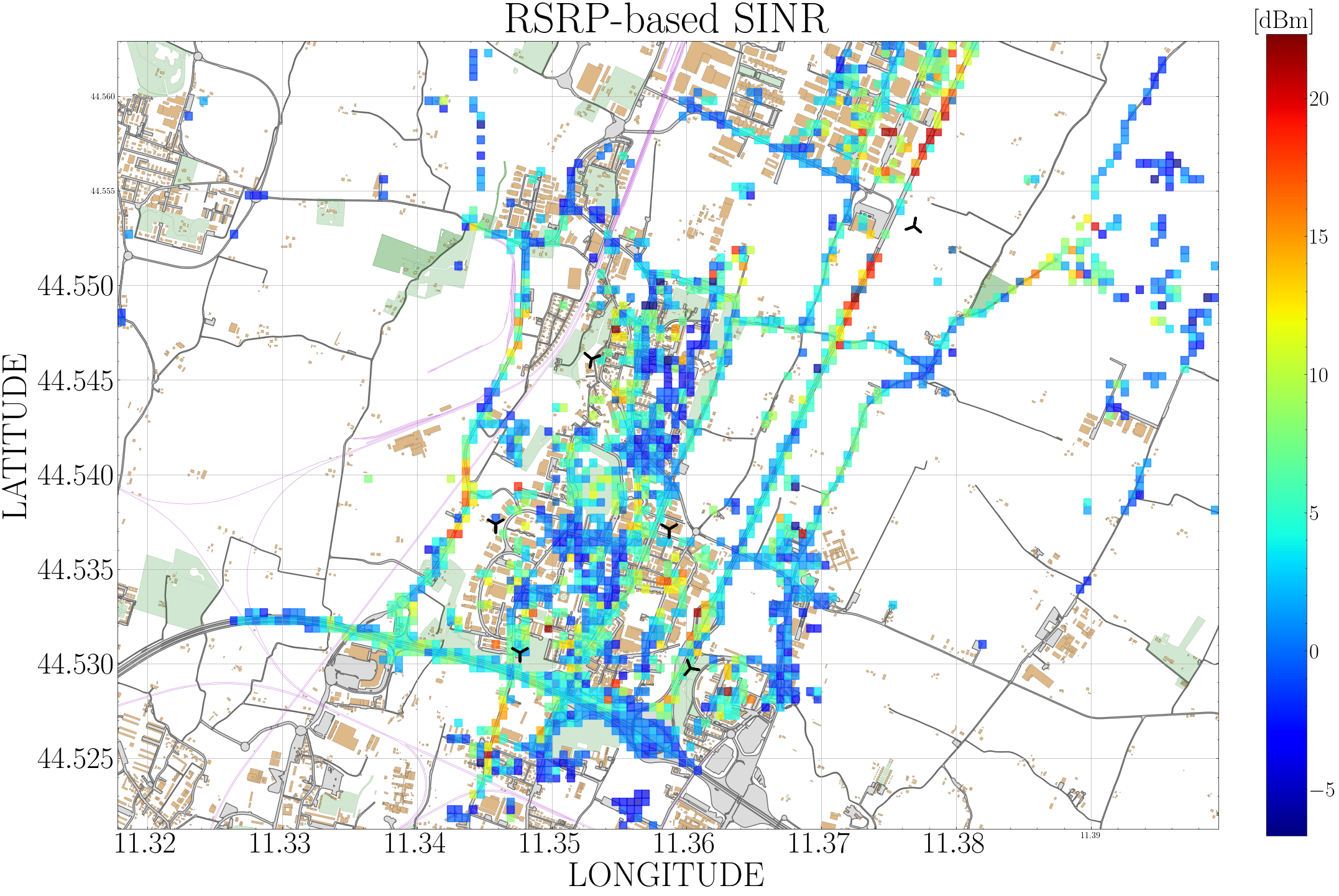}
    \caption{RSRP-based SINR}
    \label{fig:subfig_RSRP/Qb}
\end{subfigure}
\hfill
\begin{subfigure}{0.4\textwidth}
    \includegraphics[width=\textwidth]{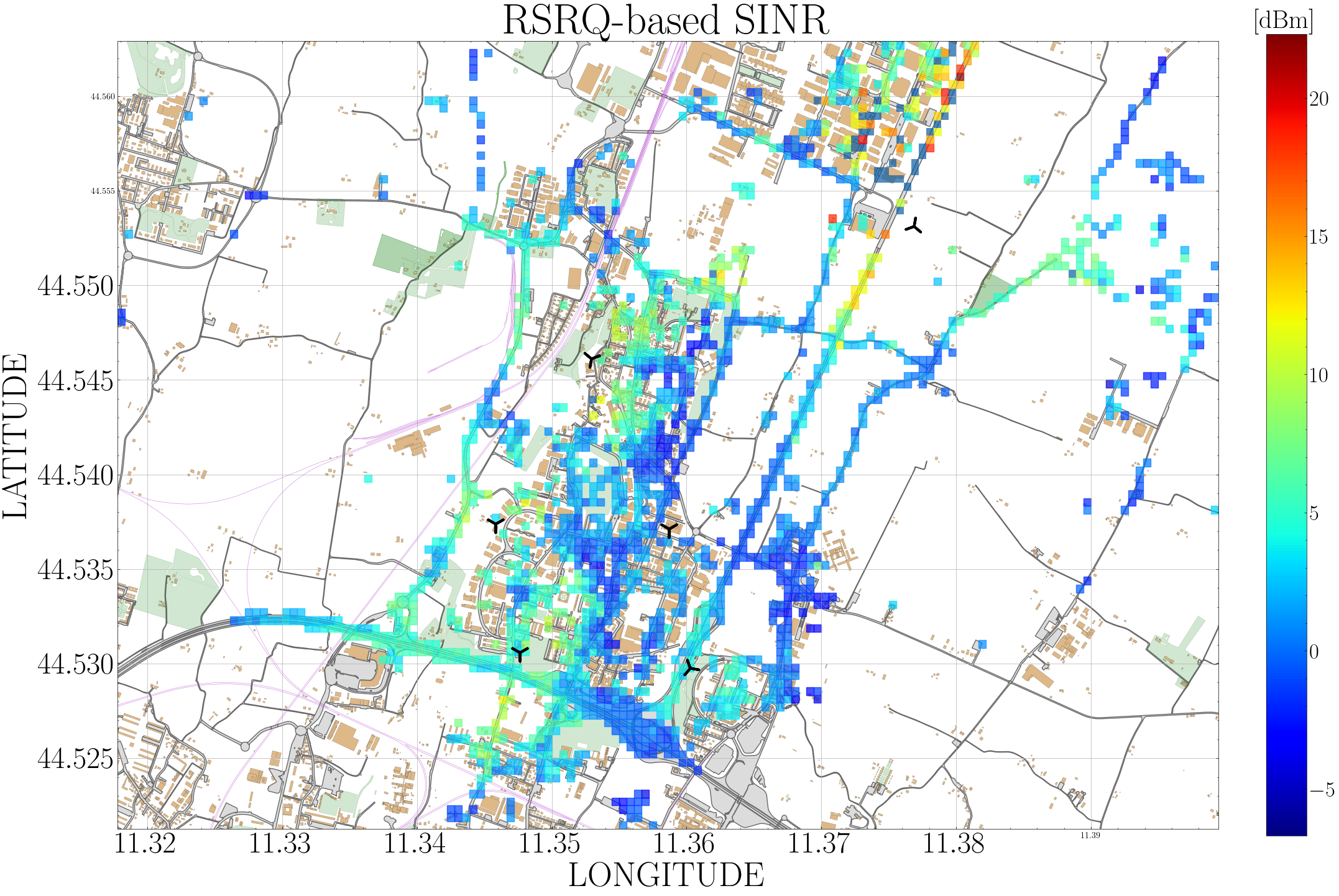}
    \caption{RSRQ-based SINR}
    \label{fig:subfig_RSRP/Qc}
\end{subfigure}
        
\caption{Comparison RSRP/RSRQ-based SINR}
\label{fig:RSRP/Q}
\end{figure*}

\subsection{Electromagnetic simulator}

Electromagnetic simulations are stored for every cell and each selectable downtilt parameter, totaling \(P\times C\) simulations, where \(P\) is the number of downtilts and \(C\) is the number of cells in the cluster. Each simulation computes the electric field intensity [dBuV/m] for a grid of 50 \([m^2] \) pixels covering the area of interest, considering a single cell as the emitting source. The electric field intensity is then converted to RSRP [dBmW] considering a reference receiver bandwidth of one Resource Element (RE) (15 [kHz]).

\noindent Once every simulation pixel’s RSRP is obtained, it is possible to compute the SINR from (\ref{eqn:(2)}). Each pixel is associated with a serving primary cell selected with a max-RSRP criterion. Ultimately, \(\Delta R\) and \(\Delta S\) are computed as the dB-difference between the simulated and MDT RSRP and SINR for every pixel. These two quantities are representatives of the electromagnetic simulation tool’s estimation error with respect to the UEs’ measurements reports. Such error is assumed to be independent of the antenna configuration on the field, meaning that the same difference would be obtained from MDT data and simulations relative to a set of antenna parameters \(A=\{P_1, P_2, ..., P_C\}\), or MDT data and simulations relative to a different set \(B=\{P’_1, P’_2, ..., P’_C \}\).

\subsection{Step method}

The step method implements an agent's action's effect on network performance, given a state observation. The immediate (or step) reward, described in Sec. 4.3., is returned as a parameter. The method entails the joint use of MDT data, em simulations, and KPIs to modify the network state. Since it is infeasible to collect MDT measurements corresponding to each agent's action, the step method defines a set of procedures to obtain synthetic MDT data (\(MDT')\) by making use of the previously computed \(\Delta\) and simulation results. Every time the step method is called, the following procedures are executed:

\begin{enumerate}
\item Weight initialization: Weight sampling is performed from a Poisson distribution, as described in the section above.
\item The number of \textit{act\_UEs} is re-distributed among the pixels based on their weight. Each pixel is associated with a \% of the total number of users described by the KPI.
\item A set of \(C\) simulations, containing all pixels' simulated RSRP, is retrieved from memory according to the cluster antennas configuration at step \(t+1\). The latter is fully determined by the agent's chosen actions from the beginning of an episode, as better specified in the next section.
\item Pixels Cell Reselection: each pixel's simulated RSRP is selected as the 
\[max[RSRP_1, RSRP_2, ... , RSRP_C]\;.\]
Each vector's element indicates the simulated RSRP from one of the \(C\) cells. The serving cell is re-assigned accordingly.
\item Each pixel's simulated SINR is re-computed as (\ref{eqn:(2)}).
\item \(MDT'\) RSRP and SINR values are obtained according to Alg. \ref{alg: step}

\begin{algorithm}
\caption{Step method}
\begin{algorithmic}
\For {i in pixels}
	\State $MDT'_{RSRP, i} =  SIM_{RSRP, i} + \Delta R_i$
	\State $MDT'_{SINR, i} =  SIM_{SINR, i} + \Delta S_i$
		  \EndFor
\end{algorithmic}
\label{alg: step}
\end{algorithm}

\item Based on newly assigned pixels (point 4), each cell's \textit{act\_UEs} is recomputed (every pixel carries a \% of active UEs).
\item Step-reward computation.
\end{enumerate}

\section{Problem formulation}

A Reinforcement Learning approach requires a formulation of the optimization problem as a Markov Decision Process, in which the action-space, state-space, and reward must be appropriately defined.

\subsection{Action Space}

\begin{figure}[ht]
\includegraphics[width=0.3\columnwidth]{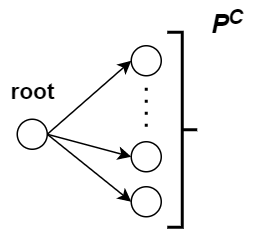}
\centering
\caption{single-step MDP}
\label{Fig: single-step MDP}
\end{figure}

\noindent The agent operates in a \(NP\)-Hard setting since it has to jointly optimize \(P\) parameters' configuration on a set of \(C\) cells. The overall number of combinations grows exponentially as \(P^C\). This formulation is represented in Fig. \ref{Fig: single-step MDP} as a single-step MDP where \(P^C\) branches generate from a root node.

\noindent Such formulation would be highly inefficient and infeasible from the computational point of view, particularly when moving to medium or large cluster sizes. Instead, it is much more convenient to represent the optimization problem with an MDP tree formulation, where the depth of the tree depends on the number of cells. At each time step of a fixed-length episode, the agent must decide 1) the selection of a target cell among the \(c\) cells and 2) a proper configuration of its \(P\) parameters. The cardinality of the action set at each time step is then reduced to \(P\times C\).

\begin{figure}[ht]
\includegraphics[width=0.9\columnwidth]{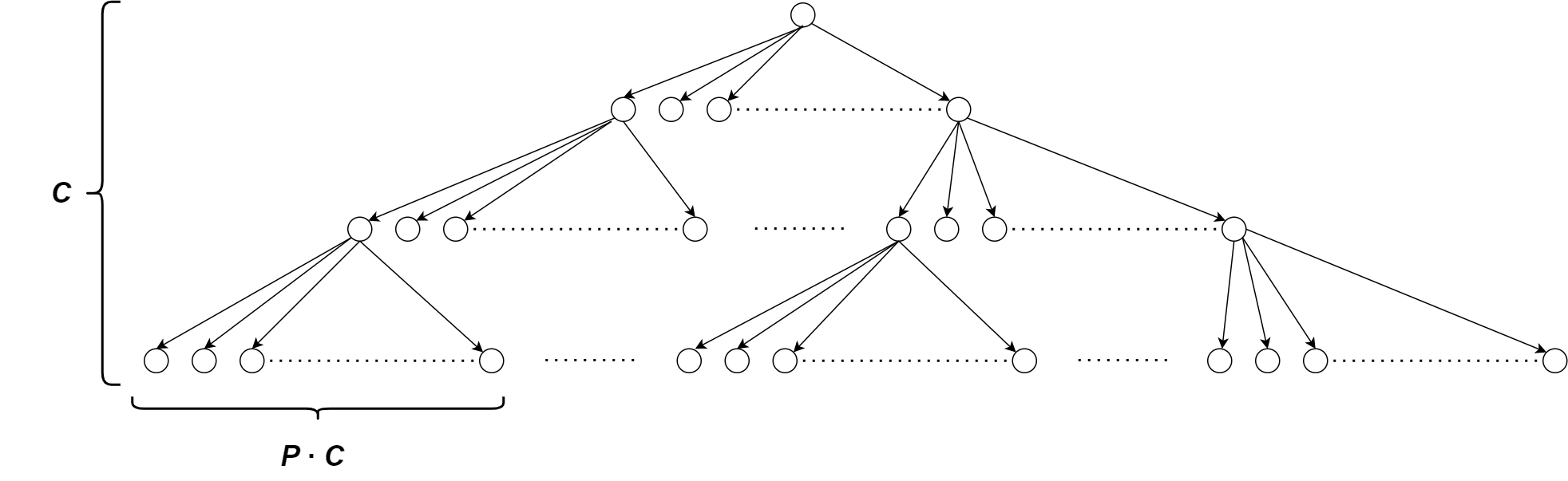}
\centering
\label{Fig: multiple-step MDP}
\caption{multiple-step MDP}
\end{figure}
\noindent Even though the overall number of nodes in a complete tree grows as \((P\times C)^{C+1}-1\), therefore faster than the previous formulation, this MDP formulation unlocks the potential to sequentially prune the tree based on the estimated Q-values at training time. Here lies the exploitation vs. exploration tradeoff: quickly exploiting the gained knowledge about the state space (Q-values estimates) translates to a faster tree pruning, as non-promising branches are discarded. Viceversa, an extended exploration phase allows the agent to explore the state space more exhaustively, reducing the risk of getting stuck into local optima.\\
To this end, the DW-\(\epsilon\eta\)-greedy exploration policy is presented in the next subsection as an improvement of the classical \(\epsilon\)-greedy policy to introduce soft constraints at training time and to tackle this tradeoff efficiently.
A set of 5 possible discrete tilts, each corresponding to an additional \(-\Delta 2\)° with respect to the ground perpendicular direction, can be selected by the agent for each of the 9 target cells of the cluster. The set of new cells’ parameters configuration is given by: \(A=\{T_1, T_2, ..., T_p\}\) at the end of each episode.

\subsection{State Space}

\noindent At each time step, the DQN agent receives as input an observation of the network state, represented by the quantities in Tab. \ref{Tab: state space}.
\begin{table*}[t]
\centering
\resizebox{0.7\textwidth}{!}{%
\begin{tabular}{|c|c|c|c|}
\hline
\textbf{RSRP} & \textbf{WEIGHT} & \textbf{SINR} & \textbf{Episode History (channel 4 $\rightarrow$ (\(P\times C +3\)))}\\
\hline
\(Y\times Z\) pixels & \(Y\times Z\) pixels & \(Y\times Z\) pixels & 1 hot-encoding bits\\
\hline
\end{tabular}%
}
\centering
\caption{State space}
\label{Tab: state space}
\end{table*}
The reference area of interest can be represented by a matrix of \(Y\times Z\) pixels composed of three different channels:
\begin{itemize}
\item RSRP is used to manage mobility and identify coverage issues.
\item WEIGHT is used to distinguish between most and less relevant pixels.
\item SINR is used for the computation of the throughput.
\end{itemize}

\noindent All channels are normalized before being processed by the agent's DNN.

\noindent Additionally, (\(P\times C\)) one-hot encoding bits representing episode history are embedded in the state space. The \(i\)-th bit, corresponding to the \(i\)-th action, will be set to 1 if action i is selected at the current step \(t\). The episode history is then reset to 0 at the beginning of each new episode. The reason behind this choice is twofold:
\begin{enumerate}
\item Memory of the past actions can be exploited.
\item A uniform replay buffer with random sampling can be employed without losing information of episodes' past actions. Although either two contiguous steps or two uncorrelated episodes' steps are sampled during mini-batch extraction, the episode history information is fully preserved at any time (Fig. \ref{fig: replay buffer}).

\begin{figure}[ht]
\centering
\includegraphics[width=\columnwidth]{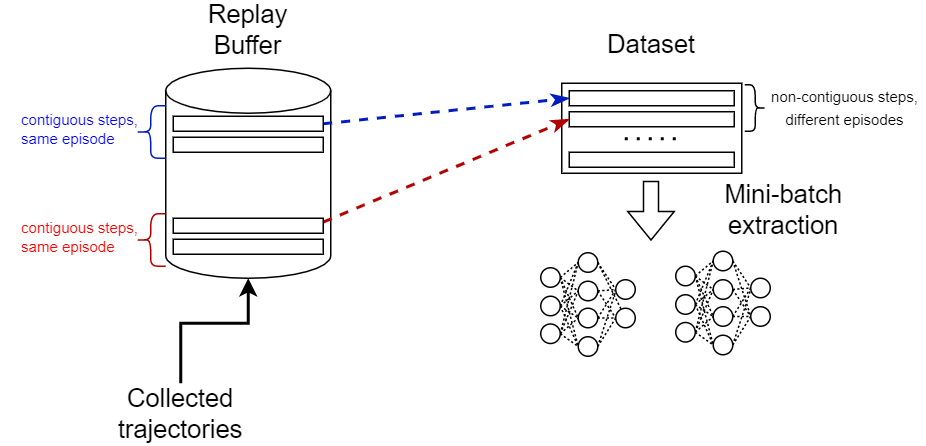}
\caption{Replay buffer and mini-batch extraction}
\label{fig: replay buffer}
\end{figure}

\end{enumerate}

\subsection{Reward}

\noindent The immediate reward function received at time step \(t+1\) after selecting action \(a_t\) must map all network performance tradeoffs in one scalar value. There is an evident twofold tradeoff of maximizing sum-rate for CCO problems without overlooking cell-edge coverage. As a consequence, two elements concur to the final reward function computation:
\begin{enumerate}
\item Average end-user throughput estimate.
\item A cost function for penalizing sets of actions violating coverage constraints.
\end{enumerate}
The cumulative future reward is the discounted sum of all step-rewards obtained during an episode, as indicated in (\ref{eqn:(1)}). Since each step corresponds to optimizing a particular cell among the ones in the considered cluster, and all cells share the same importance, the discount factor \(\gamma\) is set equal to 1. The following subsections show how the two elements are computed.

\begin{figure}[ht]
\centering
\includegraphics[width=\columnwidth]{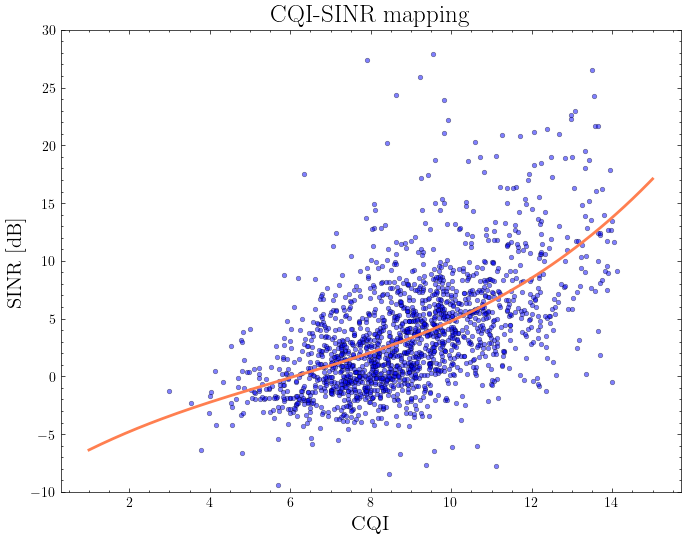}
\caption{CQI-SINR mapping from MDT data analysis}
\label{Fig: CQI-SINR}
\end{figure}

\subsubsection{Average end-user throughput estimate}

\begin{equation}
U_{\text{avg,cell}} = \eta_ {\text{MCS,cell}}\left[\frac{\text{bits}/\text{s}}{\text{Hz}}\right]\cdot N_{\text{PRB,cell}}\cdot 180[\text{KHz}]
\end{equation}
Where \(U_{avg,cell}\) is the average cell user throughput, \(\eta_ {MCS,cell}\) is the average cell spectral efficiency, \( N_{PRB,cell}\) is the average number of scheduled PRBs per UEs per cell and \(180\) [KHz] is the PRB bandwidth. \(\eta_ {MCS,cell}\) is obtained as a weighted sum of each cell's pixels spectral efficiencies:

\begin{equation}
\eta_ {MCS,cell} = \frac{1}{\sum_{i=1}^{N_{pixels,c}}w_i}\sum_{i=1}^{N_{pixels,c}}w_i \eta_{MCS,i}
\end{equation} where \(w_i\) is the pixel’s weight, as indicated in the state space subsection. Pixel’s spectral efficiency \(\eta_{MCS,i}\) is obtained according to a Channel Quality Indicator (CQI)- Modulation and Coding Scheme (MCS) mapping compliant to \cite{3GPP_36.213}, in which the maximum MCS is determined allowing a packet error rate of 10\%. The CQI is not uniquely identified by SINR, as it also depends on the implementation of the UE’s receiver: a sophisticated receiver can collect the incoming data at a lower SINR than a simpler one \cite{An_Introduction_to_LTE}. In order to derive an empirical law for the mapping between CQI and SINR, and MDT data analysis has been conducted on the reference territory. A best-fit line of polynomial degree 3 approximates the average MDT SINR values (y-axis) for different values of average CQI (x-axis) (Fig. \ref{Fig: CQI-SINR}, Tab. \ref{Tab: Table CQI-SINR}).

\begin{table}
\centering
\resizebox{\columnwidth}{!}{%
\begin{tabular}{|c|c|c|c|c|}
\hline
CQI	& SINR & Modulation & Code rate & Spectral Efficiency\\
\hline
0	&-	&-	&-	&-\\
1&	-6.4 dB&	QPSK	&0.076&	0.1524\\
2&	-4.8 dB&	QPSK	&0.19&	0.377\\
3& -3.4 dB&	QPSK	&0.44&	0.877\\
4& -2.2  dB&	16-QAM	&0.37&	1.4764\\
5&	-1.2  dB&	16-QAM	&0.48&	1.914\\
6&	-0.1  dB&	16-QAM	&0.60&	2.4064\\
7&	0.9  dB&	64-QAM	&0.46&	2.7306\\
8&	2.1  dB&	64-QAM	&0.55&	3.3222\\
9&	3.3  dB&	64-QAM	&0.65&	3.9024\\
10&	4.8  dB&	64-QAM	&0.75&	4.5234\\
11&	6.5  dB&	64-QAM	&0.85&	5.115\\
12&	8.5  dB&	256-QAM	&0.69&	5.5544\\
13&	10.9  dB&	256-QAM	&0.78&	6.2264\\
14&	13.8  dB&	256-QAM	&0.86&	6.9072\\
15&	17.1  dB&	256-QAM	&0.93&	7.4064\\
\hline
\end{tabular}%
}
\caption{CQI-SINR mapping. Empirical law from MDT data analysis}
\label{Tab: Table CQI-SINR}
\end{table}
\vspace{2ex}

\noindent The average number of PRBs per UEs per cell \( N_{PRB,cell}\) depends on the scheduling mechanism employed. In the following, two possible schemes are considered:
\begin{enumerate}
\item Round-robin scheduling
\begin{equation}
N_{PRB}=\ \frac{N_{PRB,TOT}\ -\ \text{Thr}}{N_{UE,CELL}}\,.
\end{equation}
Every UE is treated equivalently with the round-robin scheduling, regardless of its channel condition. \(N_{PRB,TOT}\) is a constant indicating the number of PRBs available for each LTE frequency band (100 PRBs for the 1800 MHz band), Thr is a safety PRB occupancy threshold (e.g., 10\%), and \(N_{UE,CELL}\) is the average number of active users by cell.
\item Fair scheduling
\begin{equation}
N_{PRB}=\ \frac{N_{PRB,TOT}\ -\ Thr}{{\sum_{i=1}^{M}\beta_iN}_{UE,CELL,i}}\;.
\end{equation}
With fair scheduling, every UE is assigned a different percentage of PRBs to balance the cell-edge performance; UEs are subdivided into M classes depending on their channel conditions (the first one corresponds to the best channel condition and the last, M-\(th\) class corresponds to the worst one). The term \(\beta_i\) varies between \(\beta_1\) and \(\beta_M\).  A user of the M-th class is assigned a number of PRBs which is M times greater than those assigned to a user of the first class.\\
\end{enumerate}

\noindent The final throughput is calculated as a linear weighted average of the average user throughput per cell. The weighting term is given by the relative number of UEs per cell:
\begin{equation}
U=\ \frac{1}{N_{TOT}}\sum_{j=1}^{c}{N_{UE,j} U_{avg,cell,j}}\;.
\end{equation}

\subsubsection{Coverage}

\noindent An exponential cost function is introduced to penalize degrading sets of actions for the coverage:
\begin{equation}
C=
    \begin{cases}
      0, & \text{if}\ 99,5\% <=\%A_{cov} <= 100\% \\
      \frac{1-4,6^{\%A_{cov}-101,1}}{11,2 \times 4,6^{(\%A_{cov} - 101,1)}}, & \text{if}\ 98\% <= \%A_{cov} < 99,5\%\\
      10, & \text{otherwhise}
    \end{cases}
\end{equation}

\begin{figure}[h]
\centering
\includegraphics[width=0.8\columnwidth]{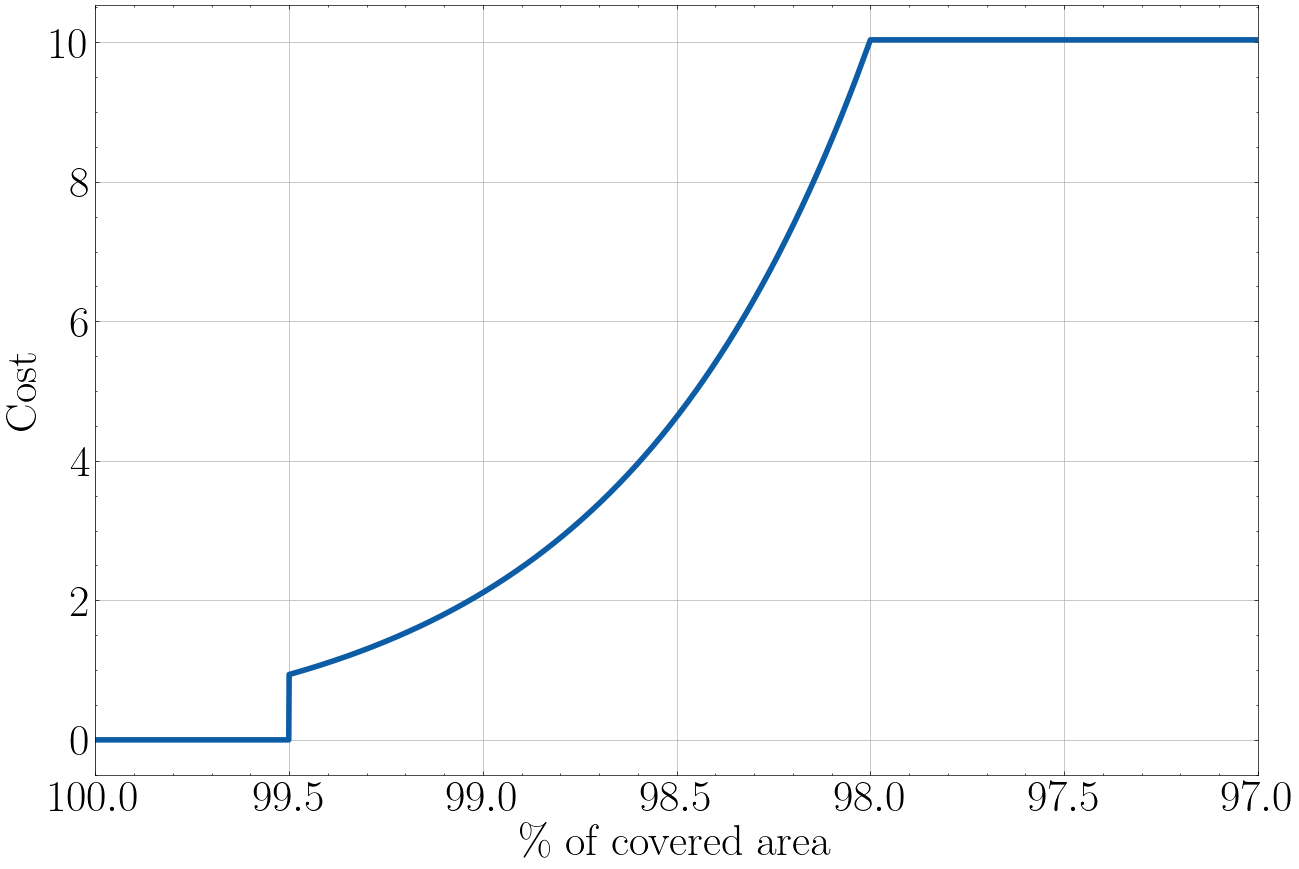}
\caption{Coverage cost function}
\label{fig: cost function coverage}
\end{figure}
The \% of covered area is evaluated as the weighted sum of the pixels in coverage \(N_{pc}\) over the total number of pixels \(N_p\), where \(N_{pc} <= N_p\):
\begin{equation}
\%A_{cov} = \frac{\sum_{i=1}^{N_{pc}} w_i}{\sum_{j=1}^{N_p} w_j}\;.
\end{equation}
As a consequence, pixels carrying more traffic are statistically more relevant. A pixel must be considered in outage if its performance are either noise limited (NL) or interference limited (IL):
\begin{enumerate}
\item RSRP < -125 dBm \(\rightarrow\) NL conditions
\item SINR < -6.4 dB (Tab. \ref{Tab: Table CQI-SINR}) \(\rightarrow\) IL conditions
\end{enumerate}

\noindent The final step-reward is the difference of the two components mentioned above:
\begin{equation}
R = U - C\,.
\label{Reward function}
\end{equation}

\section{Tweaked Deep Reinforcement Learning Agent}

\noindent This section is dedicated to the detailed description of the steps introduced in the definition of the DQN agent, aiming, in particular, to joint optimization of its sample efficiency and performance.

\subsection{Tweaking of the DQN Agent}

The agent has been realized by extending Tensorflow Agents API. The optimization steps introduced to the classical DQN formulation can be summarized into 4 points:
\begin{enumerate}
\item \textbf{Deep-tree MDP}\\ As already expressed, the first improvement consists in reformulating the action space into an episode of length \(C\), representable as a deep-tree MDP. 
\item \textbf{Episode history}\\ Introducing episode history directly into the state observation allows the agent to exploit the memory of its past actions up to the beginning of an episode.
\item \textbf{Reward shaping}\\ Constraints of the problem have been relaxed by introducing a penalty in the reward function for sets of actions that violates them. Two constraints have been introduced to the formulation:
\begin{enumerate}
\item Coverage constraint: an exponential cost function is introduced as described in the section above.
\item Optimized cells constraint: the agent receives a negative reward if it chooses to optimize a cell that has already previously modified in the same episode.
\end{enumerate}
\item \textbf{Depth-wise \(\epsilon - \eta\) greedy policy}\\ This custom collection policy applies soft constraints at training time to guide the agent's exploration phase. The proposed solution is effective in all those contexts where the agent must learn complex action patterns and events recurring with sparse probability. In particular, an extra degree of freedom is introduced with respect to the classical \(\epsilon\)-greedy policy: the \(\eta\) parameter. This parameter controls the probability of performing a "constrained random action", that is, a random action sampled from a constraint-compliant set. The pseudo-code at every episode step is presented in Algorithm \ref{exploration policy} \\

\begin{algorithm}
\caption{depth-wise \(\epsilon-\eta\) greedy policy}
\begin{algorithmic}
\State $i \gets$ random(0,1)
\State $j \gets$ random(0,1)
\State $\epsilon \gets \epsilon$\_scheduling
\State $\eta \gets \eta$\_scheduling
\If {i $\geq \epsilon$}:
	\State perform greedy action
\Else
\If {j $\geq \eta$}:
\State perform constrained random action
\Else
	\State perform random action
	\EndIf
	\EndIf
\end{algorithmic}
\label{exploration policy}
\end{algorithm}

As for the problem at hand, the desired behavior that the agent should learn is to optimize a different cell at every step of the episode. Hence, constraint (b) is enforced with probability \(\eta\). The probability of randomly playing a constraint-compliant episode is given by:
\begin{align}
P_{episode}&=P(step_{C}|step_{C-1},…,step_1 )\nonumber \\&= \prod_{i=0}^{C-1}(1-\frac{i}{C})=\frac{(C-1)!}{C^{C-1}}\,.
\end{align}
Again, \(C\) indicates the number of target cells. Since \((C-1)!=o(C^{C-1})\), the inverse of the probability grows exponentially as the size of the problem grows.
It is observable that the behavior is sparse, since \(C^{C-1}/(C-1)! \approx 1067\), for C=9. This means that exploring the space randomly (i.e., with an \(\epsilon\)-greedy policy) is highly inefficient, as only one episode out of 1067 on average is concluded without penalties. By controlling the probability \(\eta\) of choosing an action over a limited set, the number of positive experiences and penalties is balanced, and the problem of sparse positive reward is overcome. It is crucial to notice that this policy does not entirely preclude the agent from choosing an action that violates a constraint (unless \(eta=1\)). Instead, the agent might find that violating a constraint provides a better long-term reward despite the immediate negative penalty. Indeed, it is a good practice not to strictly limit the agent's action space, as the programmer does not know a priori the best solution. This is particularly true if multiple constraints are applied simultaneously or the number of episodes steps is not set to a fixed number, which is left for future work.
The benefit of the introduction of the \(\eta\) parameter, as presented in Sec. 6, is twofold:
\begin{enumerate}
\item Exploration phase time is drastically reduced.
\item Training is strongly stabilized.
\end{enumerate}
The introduction of the \(\eta\) parameter is fundamental to guide the agent in acquiring complex action patterns, but is not enough for the learning process. Indeed, the advantage of a tree formulation is that of pruning the tree based on the estimated Q-values. This allows the agent to focus only on promising branches, discarding a vast portion of the state space. Since the Q-values estimate accuracy improves with the number of training episodes, the \(\epsilon\) is usually scheduled to decrease monotonically during training (Fig. (\ref{fig: espilon scheduling}a)). Consequently, exploration leaves room for exploitation as the agent interacts and gains knowledge of the environment.
Since the number of nodes grows exponentially with the depth of the tree, the first-level nodes are visited much more often than the leaf nodes. Hence, a unique scheduling of the \(\epsilon\) parameter is not the most effective solution. Exploring the first-level nodes for a shorter period, instead, would be much more convenient in order to allow the tree to be pruned sequentially. To this end, a depth-wise scheduling of the \(\epsilon\) parameter is proposed: at each episode step, the \(\epsilon\) probability is sampled from a different scheduling function. The first level scheduling decays faster than the last layers' one (Fig. (\ref{fig: espilon scheduling}b)).

\begin{figure}[h]
\centering
\begin{subfigure}{\columnwidth}
    \includegraphics[width=\textwidth]{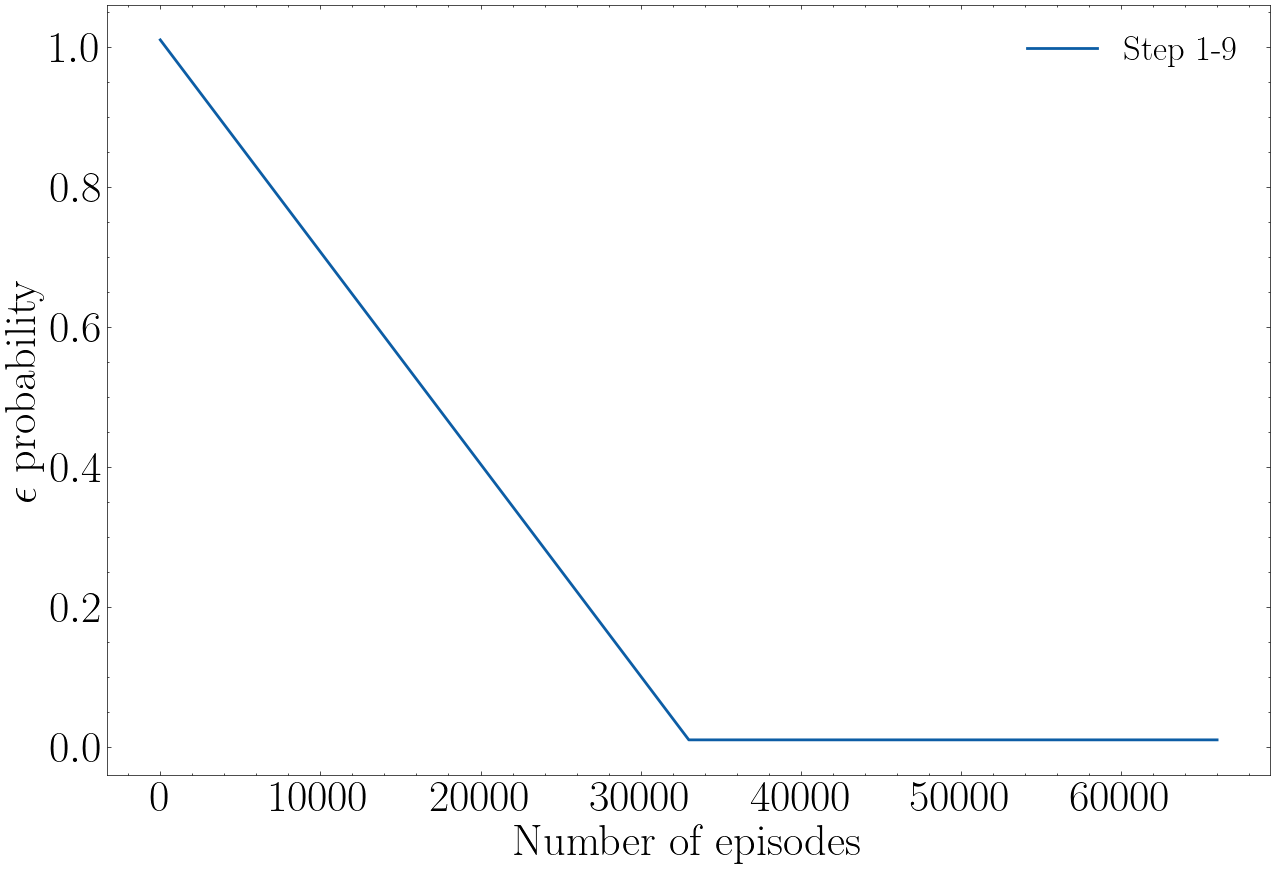}
    \caption{Monotonically decaying epsilon scheduling}
    \label{fig:subfig_scheduling_a}
\end{subfigure}
\hfill
\begin{subfigure}{\columnwidth}
    \includegraphics[width=\textwidth]{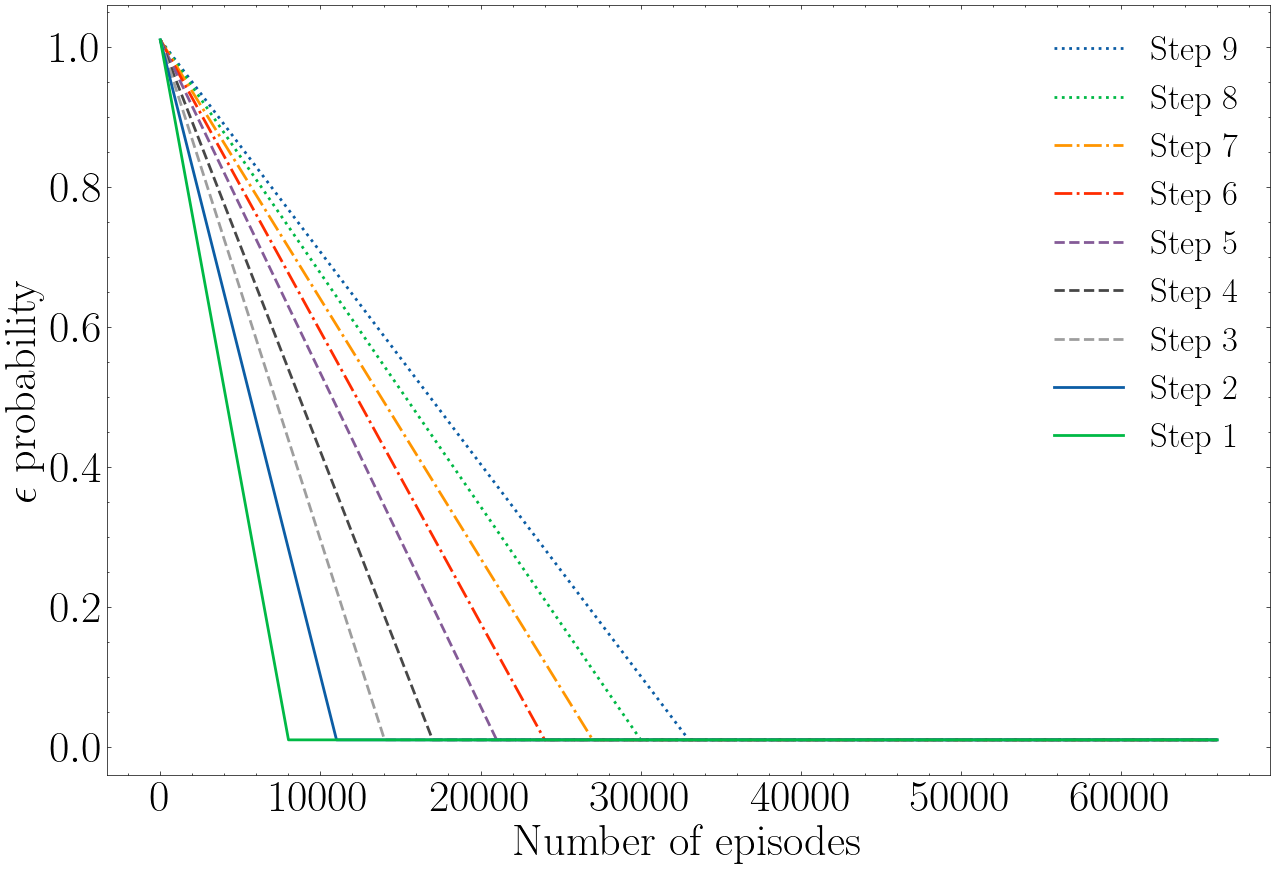}
    \caption{Depth-wise epsilon scheduling}
    \label{fig:subfig_scheduling_b}
\end{subfigure}
\caption{Scheduling of the \(\epsilon\) parameter.}
\label{fig: espilon scheduling}

\end{figure}

This leads to sequential pruning of the tree, much more efficient in terms of convergence time and quality of the found solutions.\\

The parameter \(\eta\) can be scheduled according to the episode's step too. Since the probability that the desired behavior randomly occurs is lower at the end of the episode (more cells have already been optimized), it makes sense to have higher \(\eta\) values for more profound steps. In particular, a desired behavior would be to spread evenly among the tree levels the overall number of positive and negative experiences. This can be computed in terms of probabilities:

\begin{equation}
    P_{s,step} = P_{s}(\bar{\eta})(1-\eta) + P_{s}(\eta)\eta = P_{s}(\bar{\eta})(1-\eta) + \eta
\end{equation}
\begin{equation}
    P_{s,ep} = P(s_{step,C} | s_{step,C-1}, ..., s_{step,1}) = \prod_{i=1}^{C}P_{s,i}
\end{equation}
where \( P_{s,step}\) denotes the probability of respecting constraint (b) for a single episode step, \(P_{s,ep}\) the probability of respecting it for over the whole length of the episode, and \(\bar{\eta}\) is the complementary probability of \(\eta\). Setting a proper target balance between successful and unsuccessful episodes, e.g., \(P_{s,ep}\) = 0.5, assuming i.i.d episodes steps, it is easy to obtain that \( P_{s,i} = P_{s,ep}^{(1/C-1)}\). For \(C=9\) and \(P_{s,ep}=0.5\), we obtain \(P_{s,i} \approx 0.917\). From this, we can derive the \(\eta\) values for every episode step:
\begin{equation}
    \eta = \frac{P_{s,i} - P_{s}(\bar{\eta})}{1-P_s(\bar{\eta})}
\end{equation}
\begin{table}[t]
    \centering
    \resizebox{.275\textwidth}{!}{%
    \begin{tabular}{|c|c|c|}
    \hline
    \textbf{Episode step} & \textbf{\(P_s(\bar{\eta})\)} & \textbf{\(\eta\)}\\
    \hline
    1 & 1 & 0\\
    2 & 8/9 & 0.253\\
    3 & 7/9 & 0.627\\
    4 & 6/9 & 0.751\\
    5 & 5/9 & 0.813\\
    6 & 4/9 & 0.851\\
    7 & 3/9 & 0.876\\
    8 & 2/9 & 0.893\\
    9 & 1/9 & 0.907\\
    \hline
    \end{tabular}%
    }
    \caption{depth-wise \(\eta\) scheduling}
    \label{tab:my_label}
\end{table}

Therefore, unlike \(\epsilon\), \(\eta\) does not decay with the number of episodes but remains constant throughout the training (Tab. \ref{tab:my_label}).

\end{enumerate}

\subsection{DQN Architecture}

\begin{figure*}[t]	
\centering
\includegraphics[width=0.7\textwidth]{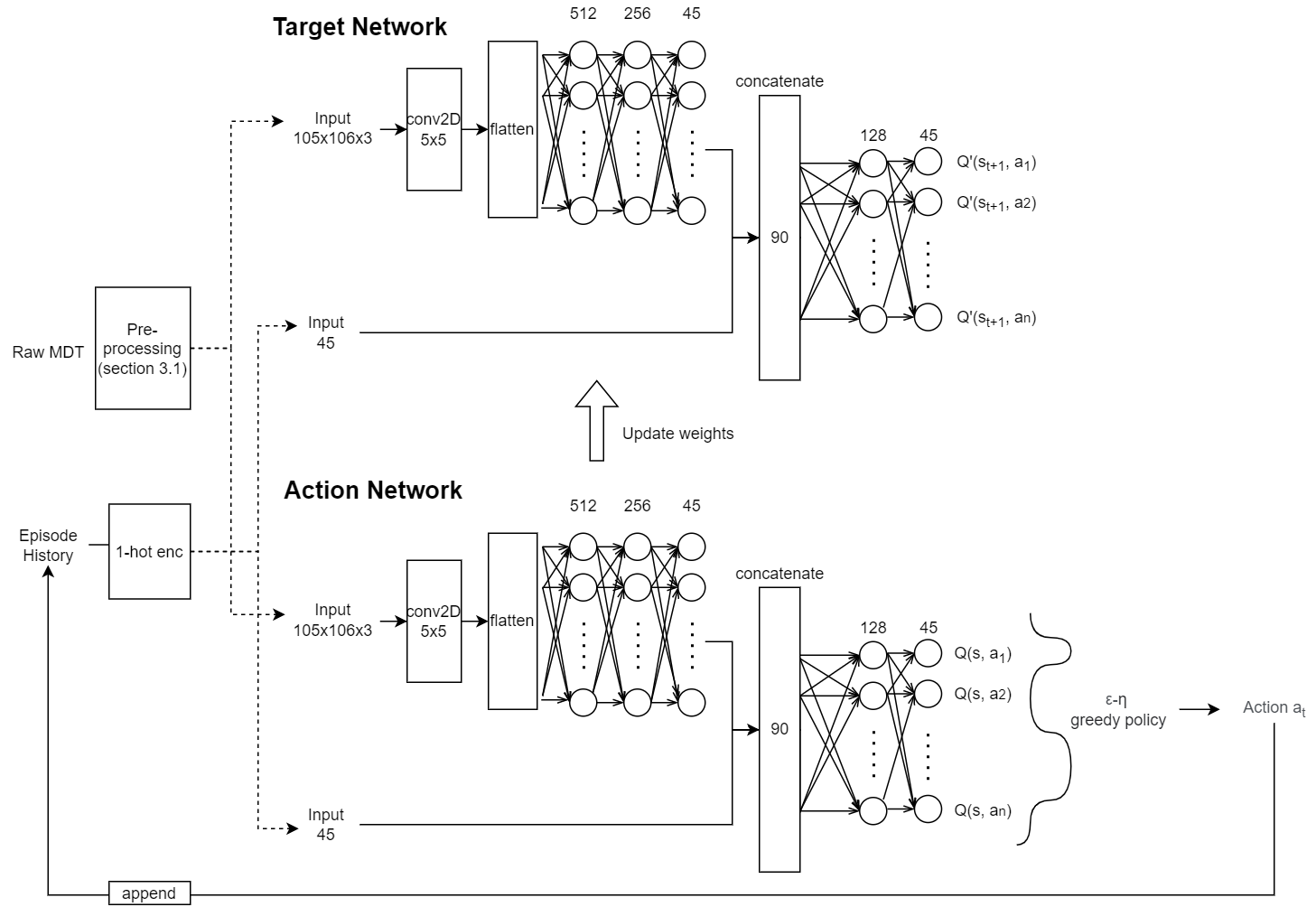}
\caption{Deep Q-Network architecture}
\label{fig:DQN_architecture}
\end{figure*}

The DQN architecture employed for the training is depicted in Fig. (\ref{fig:DQN_architecture}).

\noindent Input data are distinguished between MDT pixel at the output of the pre-processing block (Sec 3.1) and episode history. Two distinct branches of the DQN process the two data streams before being concatenated and fed to one common fully-connected layer. The former is processed by one 5x5 2D Conv layer and three fully connected layers. No pooling layers are employed since we do not want to introduce translational invariance: pixels' location matters. The latter, instead, go through a pass layer directly into the concatenate block. A detail of the chosen hyperparameters is provided in Tab. \ref{tab: Hyperparameters}. A fixed Q-targets variant of the algorithm is utilized, employing two DQNs as target and action networks with an update frequency of 1500 training steps (Fig. (\ref{fig:DQN_architecture})).

\quad
\begin{table}[t]
    \centering
    \resizebox{\columnwidth}{!}{%
    \begin{tabular}{|c|c|c|c|}
    \hline
    \textbf{Hyperparameter} & \textbf{Value} & \textbf{Hyperparameter} & \textbf{Value} \\
    \hline
    Replay buffer size & \(2\times10^5\) & Conv2D layer & 1, 5x5 \\
    Batch size & 256 & Stride & 1 \\
    Epochs & 66000 \(\times\) 9 & Learning rate & 2,5\(\times10^{-4}\)\\
    Optimizer & RMSprop & RMSprop's rho & 0.95\\
    Target update freq & 1500 & Momentum & 0.0\\
    Discount factor & 1.0 & Activation function & ReLU \\
    \hline
    \end{tabular}%
    }
    \caption{Hyperparameters table}
    \label{tab: Hyperparameters}
\end{table}

\section{Observed Results}

\noindent The proposed tweaked DQN algorithm has been trained for a total of 66000 episodes, or 594000 steps. The collected results show performance improvements in terms of 1) quality of the found solution (episode reward), 2) sample efficiency and 3) training stability with respect to few baseline algorithms: Best First Search (BFS) and DQN with \(\epsilon\)-greedy exploration policy.\\
Best First Search is a graph exploration algorithm in which each branch of the tree is exhaustively explored up to the one-level deeper node. In this way, each branch at episode step \(n\) is explored with a brute force approach up until episode step \(n+1\). Hence, the best branch is selected as source node for the next iteration. As a consequence, the algorithm selects for every episode step the action corresponding to the maximum immediate reward.\\

\noindent We present several results related to two reward functions:
\begin{itemize}
\item Case 1: The coverage-constrained user throughput reward function (Eqn. (\ref{Reward function})) expressed in Sec.4. In this case, the reward is also indicative of the gain in [Mbps] with respect to the average throughput corresponding to the default antenna configuration.
\item Case 2: Weighted average sum of MDT pixels' RSRP and SINR:
    \begin{equation}
        R = \frac{1}{\sum_i w_i}\sum_{i=1}^{N_{pixels}}w_i(RSRP_i + SINR_i).
    \end{equation}
\end{itemize}
\noindent Reward shaping with constraint (b) is applied to both reward functions for the DW-\(\epsilon\eta\)-greedy policy.\\

\begin{figure*}[h]
\centering
\begin{subfigure}{0.45\textwidth}
    \includegraphics[width=\textwidth]{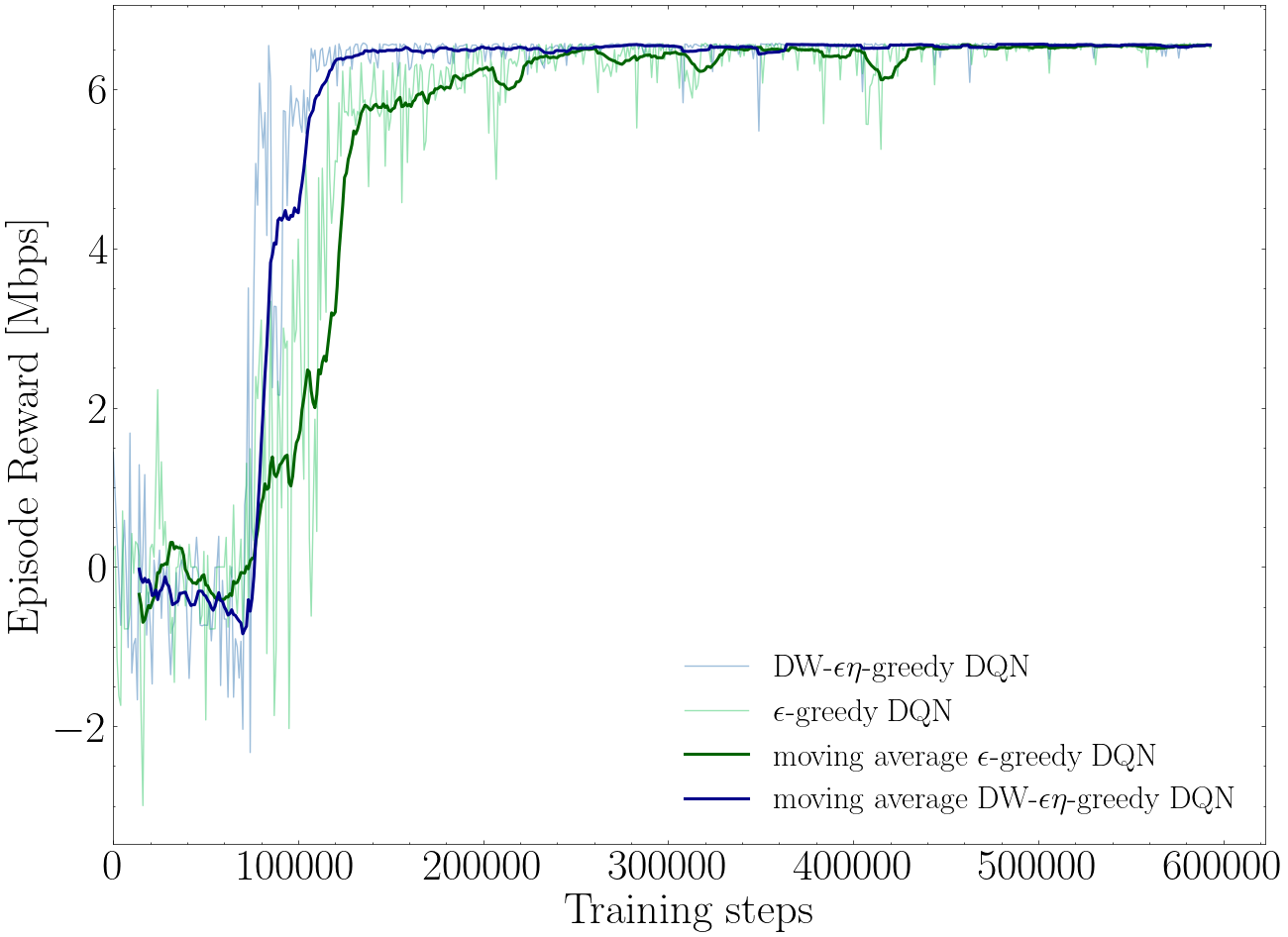}
    \caption{Reward: Coverage constrained throughput}
    \label{fig:subfig_training_a}
\end{subfigure}
\hfill
\begin{subfigure}{0.45\textwidth}
    \includegraphics[width=\textwidth]{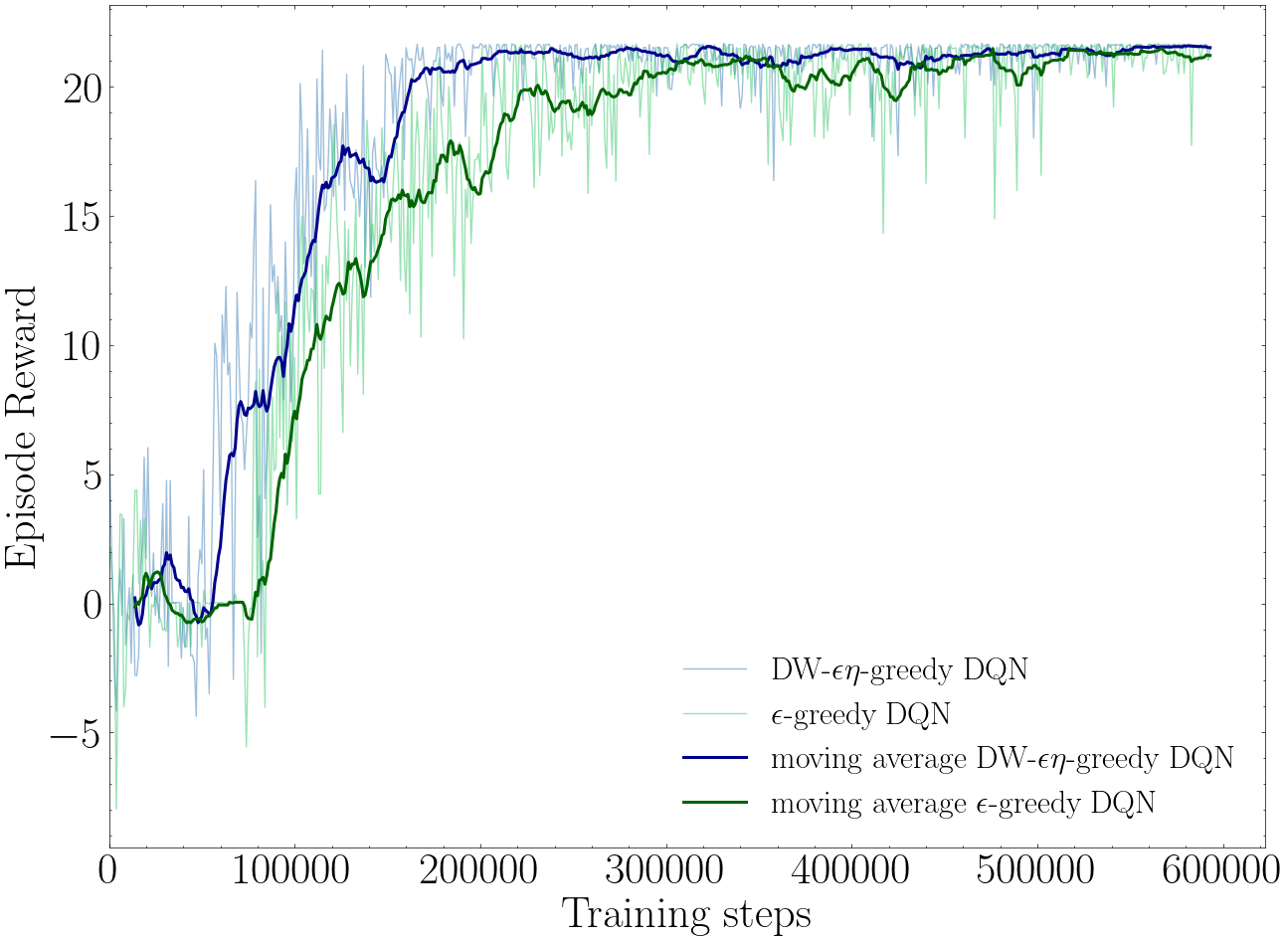}
    \caption{Reward: Average weighted RSRP + SINR}
    \label{fig:subfig_training_b}
\end{subfigure}
\caption{Training curves: DW-\(\epsilon\eta\) -greedy DQN vs \(\epsilon\) -greedy DQN}
\label{fig: DQN_vs_DQN-eta}
\end{figure*}

\noindent Fig. (\ref{fig: DQN_vs_DQN-eta}) shows the training curve of the proposed DW-\(\epsilon\eta\)-greedy DQN compared to a classical DQN with \(\epsilon\)-greedy exploration policy. The curves have been obtained by testing the two algorithms' greedy policies over 10 random seeds every 1000 episode steps during training. Additionally, their relative moving averages have been computed with a window length equal to 15 samples. Both algorithms have been trained with the same hyperparameters of Tab. \ref{tab: Hyperparameters}. Both Fig. (\ref{fig: DQN_vs_DQN-eta}a) and (\ref{fig: DQN_vs_DQN-eta}b) show that the proposed DW-\(\epsilon\eta\) -greedy DQN registers remarkable performance improvement with respect to the \(\epsilon\)-greedy DQN. In particular, the former manifest greater stability, slightly increased performance and, most importanltly, better sample efficiency, as it reaches a stable plateau in an up to 70\% shorter time (Fig. (\ref{fig: DQN_vs_DQN-eta}a)).\\

\begin{figure*}[h]
\centering
\begin{subfigure}{0.45\textwidth}
    \includegraphics[width=\textwidth]{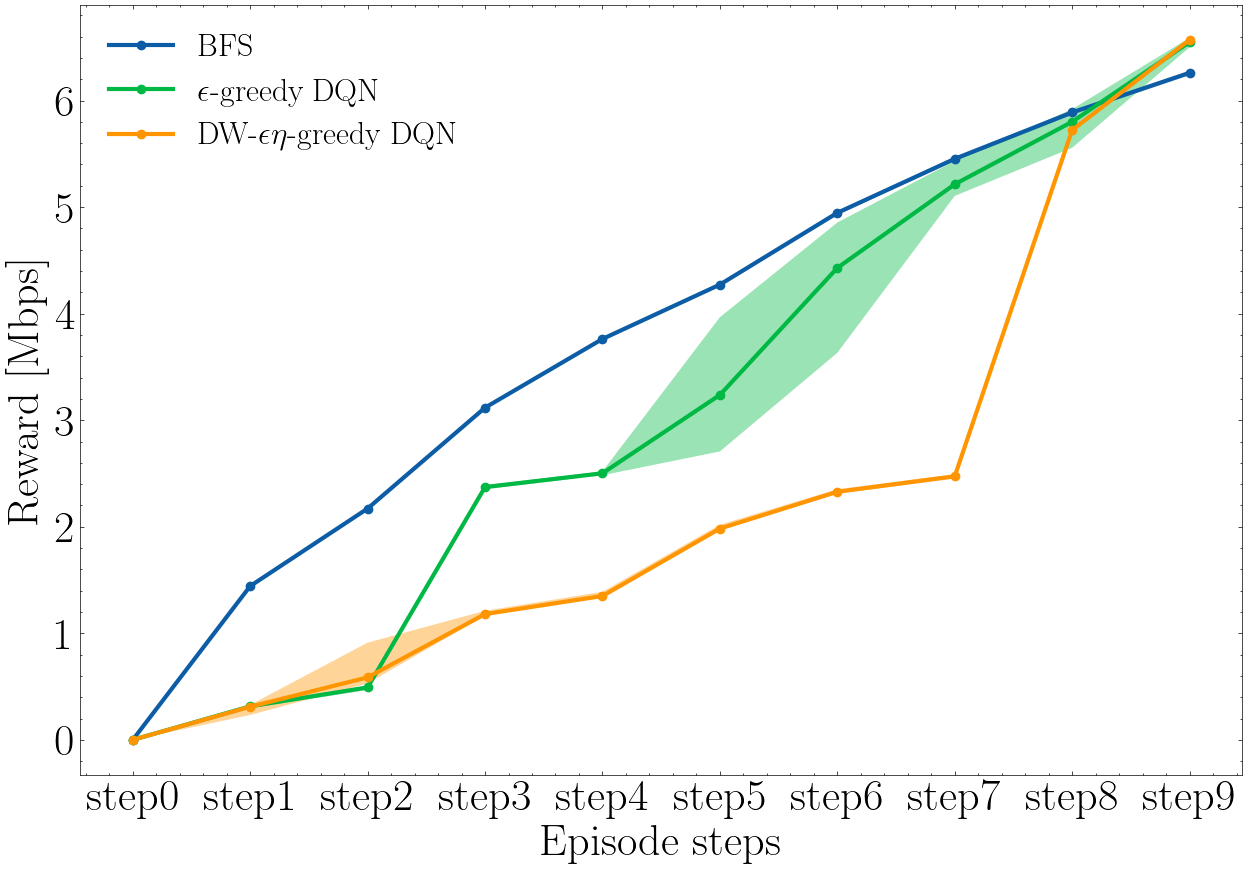}
    \caption{Reward: Coverage constrained throughput}
    \label{fig:subfig_episode_a}
\end{subfigure}
\hfill
\begin{subfigure}{0.45\textwidth}
    \includegraphics[width=\textwidth]{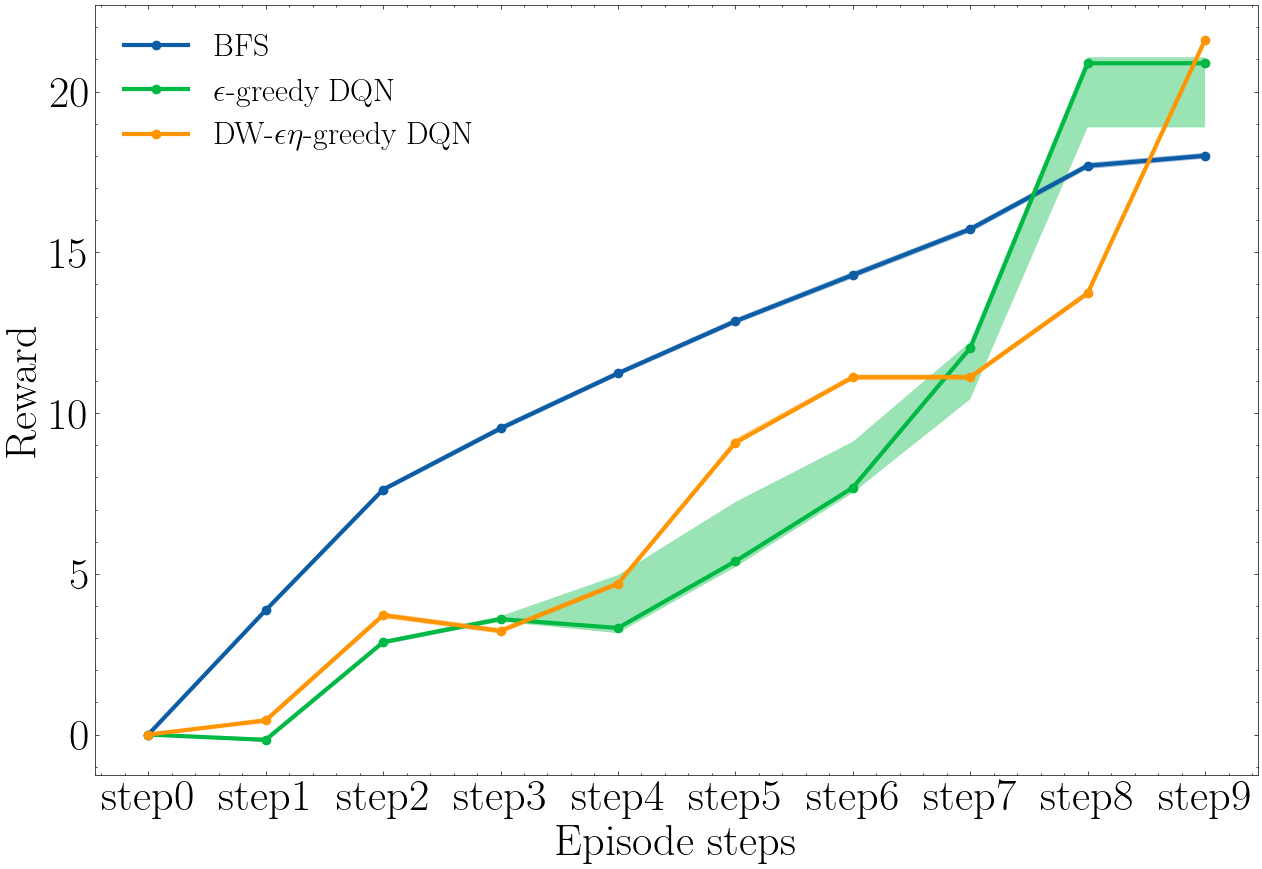}
    \caption{Reward: average weighted RSRP + SINR}
    \label{fig:subfig_episode_b}
\end{subfigure}
\caption{Step reward: DW-\(\epsilon\eta\) -greedy DQN vs \(\epsilon\) -greedy DQN vs BFS.}
\label{fig: BFS_DQN}

\end{figure*}

\noindent Fig. (\ref{fig: BFS_DQN}) depicts the performance gain obtained by our proposed algorithm with respect to BFS. The two Fig. (\ref{fig: BFS_DQN}a) and (\ref{fig: BFS_DQN}b) illustrate the reward obtained at every episode step by the three algorithms. Results are depicted togheter with their relative 99\% confidence intervals. It is particularly evident how both  \(\epsilon\)-greedy DQN and DW-\(\epsilon\eta\) -greedy DQN agents are able to learn to efficiently sacrifice the immediate reward in favour of a better cumulative episode reward (the reward value obtained at step 9, i.e., at the end of the episode). Another interesting behavior worth noticing is that the DW-\(\epsilon\eta\) -greedy DQN's policy shows more compactness than the \(\epsilon\)-greedy DQN's one: since the confidence intervals of the former are narrower, this means that more environment states map to the same antenna configurations. This translates to a less frequent need of reconfiguration for the network parameters.\\

\begin{figure*}[h]
\centering
\begin{subfigure}{0.45\textwidth}
    \includegraphics[width=\textwidth]{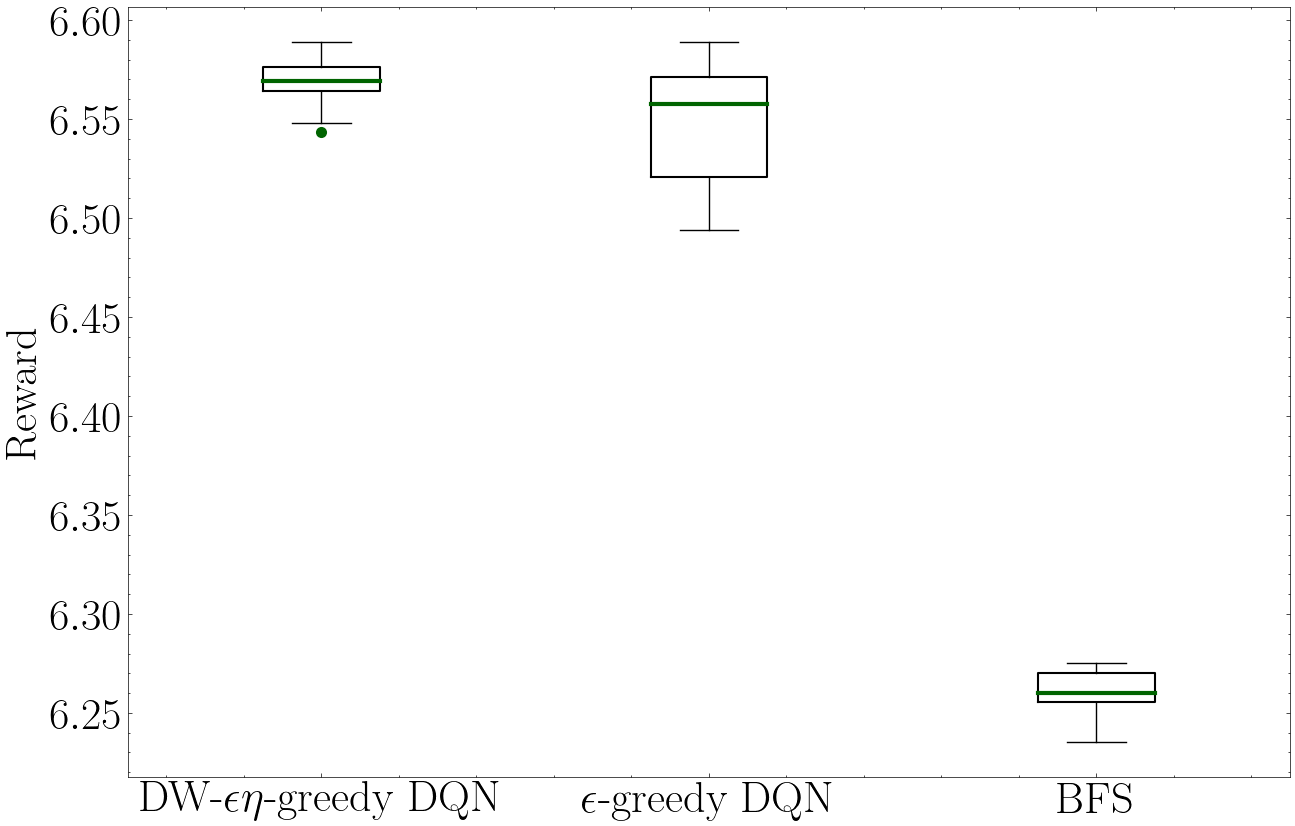}
    \caption{Reward: Coverage constrained throughput}
    \label{fig:subfig_boxplot_a}
\end{subfigure}
\hfill
\begin{subfigure}{0.45\textwidth}
    \includegraphics[width=\textwidth]{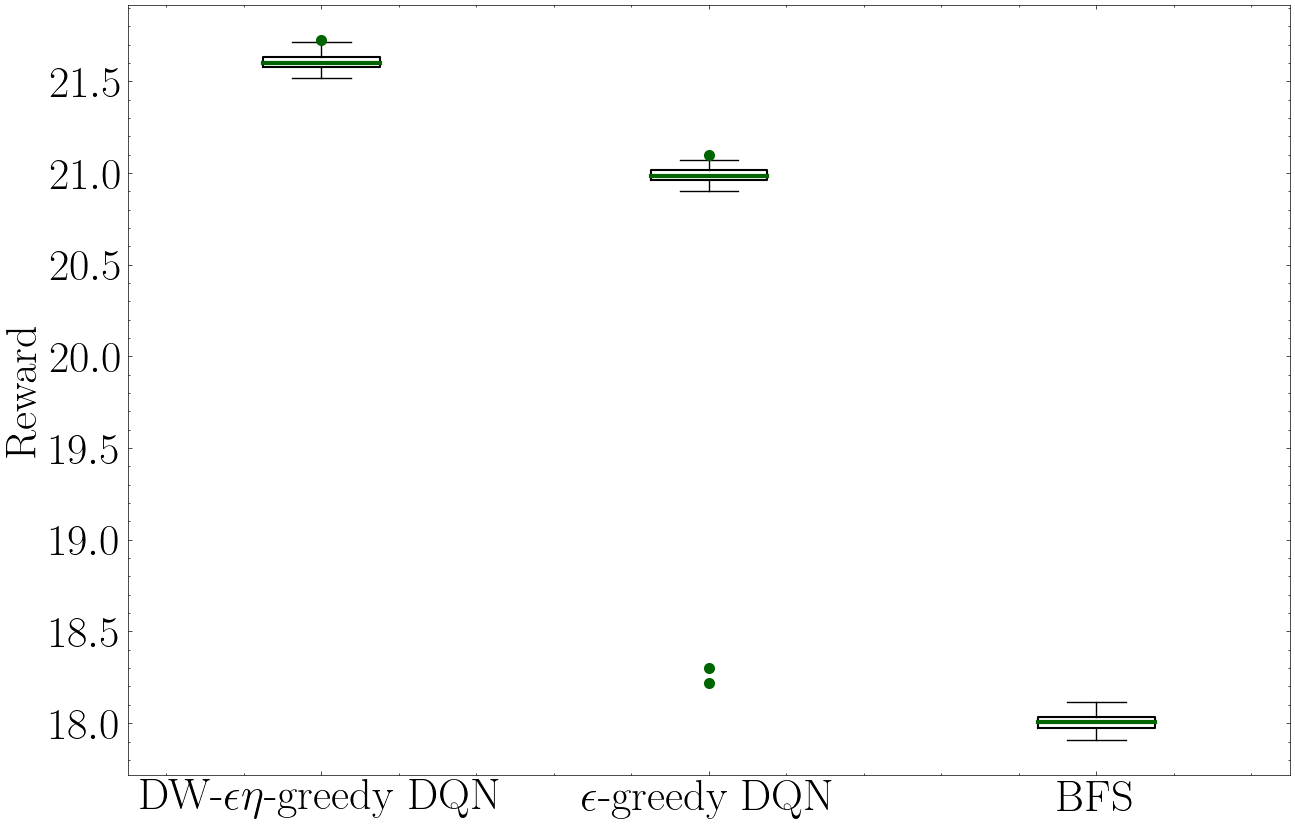}
    \caption{Reward: average weighted RSRP + SINR}
    \label{fig:subfig_boxplot_b}
\end{subfigure}
\caption{Average episode reward boxplot distribution.}
\label{fig: Boxplot_BFS_DQN}
\end{figure*}

\noindent In RL, the variance between runs is typically enough to create statistically different distributions just from varying random seeds \cite{DRL_that_matters}. As such, sensibility to random seeds is a crucial aspect of RL that needs to be addressed when assessing an algorithm's performance. Boxplots of Fig. (\ref{fig: Boxplot_BFS_DQN}) show episode reward distribution for the three algorithms when tested over 50 different random seeds. DW-\(\epsilon\eta\) -greedy DQN not only outperforms the baselines algorithms in terms of average episode reward, but shows a variance value comparable to that of BFS.

\begin{figure}[h]
\centering
\includegraphics[width=\columnwidth]{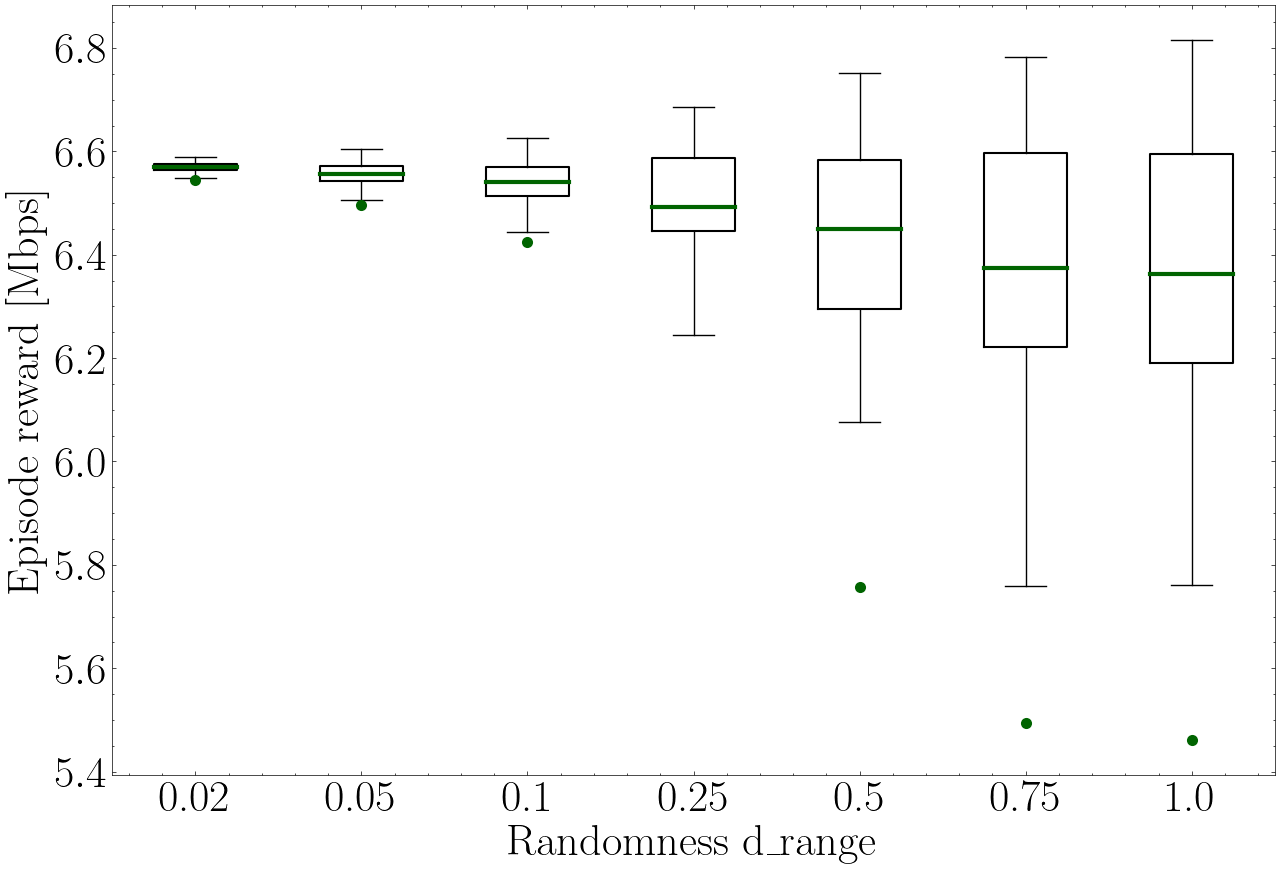}
\caption{DW-\(\epsilon\eta\) -greedy DQN performance with increasing randomness in traffic distributions}
\label{fig: stress_test_d_range}
\end{figure}

\noindent Finally, Fig. (\ref{fig: stress_test_d_range}) evaluates the algorithm's stability for increasing level of randomness in the environment's traffic distribution. In order to simulate such randomness, multiple runs over 50 random seeds have been executed. In each run, the average pixels' WEIGHT is varied every episode according to a uniform distribution with increasingly larger boundaries:
\begin{equation}
w_i^{'} = w_{i,default} + U[- d_{range}/2, + d_{range}/2] \;
\end{equation}
Where \(U[a,b]\) indicates the uniform distribution between \(a\) and \(b\). After all \(w_i^{'}\) values are re-computed, the weights are then normalized in the range [0,1].

\noindent As expected, results show larger variance for increasing values of \(d_{range}\). Nevertheless, the agent shows substantial stability, never experiencing rewards below +5.5 [Mbps] even in the worst case scenario.

\section{Conclusion}

In this work, we address a CCO problem in a real network scenario using an tweaked MDT-driven DRL agent to select optimal antenna downtilts configuration for a cluster of cells of TIM's network. In particular, we leveraged the joint use of MDT data, network KPIs and electromagnetic simulations to realize an accurate simulated network environment for the training of a tweaked DRL agent. The simulated network environment is able to accurately represent the effect of the change of a configuration parameter (e.g., antenna downtilts) on network performance. Furthermore, an optimized version of the DQN formulation with a custom exploration policy has been introduced. The obtained results show remarkable improvements in terms of sample efficiency, stability, and quality of found solution with respect to baseline algorithms like BFS and DQN with \(\epsilon\)-greedy exploration policy.

\noindent Future work might focus on the joint optimization of antenna tilts and other network parameters, as well as on a multi-agent approach to target the optimization problem of larger or denser geographical areas.

\section{Patents}

A patent application resulting from the work reported in this manuscript has been filed with the title “Optimization of the configuration of a mobile communications network”. The same authors of the manuscript are involved in the patent application.

\bibliographystyle{unsrtnat}
\bibliography{main}

\end{document}